\documentclass[aps,jcp,superscriptaddress,amsmath,amssymb]{revtex4-1}

\usepackage{graphicx}% Include figure files
\usepackage{dcolumn}% Align table columns on decimal point

\usepackage{bm}% bold math
\usepackage[english]{babel}

\usepackage[dvipsnames]{xcolor}

\usepackage[utf8]{inputenc}
\usepackage[T1]{fontenc}
\usepackage{mathtools}
\usepackage{placeins}

\usepackage{lipsum}

\newcommand{\rref}[1]{Eq.~(\ref{#1})}
\newcommand{\rrefsa}[1]{Eqs.~(\ref{#1})}
\newcommand{\rrefsb}[1]{(\ref{#1})}
\newcommand{\bA}{\boldsymbol{A}}

\newcommand{\bs}{\boldsymbol{s}}

\newcommand{\br}{{\boldsymbol{r}}}

\newcommand{\bI}{\boldsymbol{I}}

\newcommand{\bE}{\boldsymbol{E}}

\usepackage{dcolumn}%
\newcolumntype{.}{D{.}{.}{8}}

\DeclareUnicodeCharacter{2009}{\,}
\usepackage{bm}
\usepackage{mathtools}
\usepackage{physics}
\usepackage{comment}

\newcommand{\bos}[1]{\boldsymbol{#1}}
\newcommand{\bR}{{\bf R}}
\newcommand{\mr}[1]{\mathrm{#1}}
\def\iim{\mr{i}}
\def\eem{\mr{e}}

\def\nb{N_\text{b}} 
\def\DC{\text{DC}}
\def\proj{\text{proj}}

\def\Eh{\text{E}_\text{h}}

\def\balpha{\bos{\alpha}}

\def\unitfour{1^{[4]}} %{1_4}
\def\tT{\text{T}}
\def\nnuc{N_\text{nuc}}

\newcommand{\oo}[1]{\mathcal{O}(\alpha^{#1})}

\usepackage{color}
\usepackage{multirow}

\usepackage[unicode]{hyperref}
\hypersetup{
   unicode=true,          % non-Latin characters in Acrobat??s bookmarks
   plainpages=false,
   colorlinks=true,       % false: boxed links; true: colored links
   citecolor=blue,        % color of links to bibliography
}

\usepackage{url}

\usepackage{tikz}
\usetikzlibrary{trees}
\usetikzlibrary{decorations.pathmorphing}
\usetikzlibrary{decorations.markings}
\tikzset{
    photon/.style={decorate, decoration={snake,segment length=1.5mm}, draw=black},
    electron/.style={draw=black, postaction={decorate},
        decoration={markings,mark=at position .55 with {\arrow[draw=black]{>}}}}, 
    gluon/.style={decorate, draw=magenta,
        decoration={coil,amplitude=4pt, segment length=5pt}},
    boundelectron/.style={thick, double},
    coulomb/.style={dashed,thick}
}
\DeclareUnicodeCharacter{2212}{-}

\usepackage{soul}

\begin{document}

\title{Variational Dirac--Coulomb explicitly correlated computations for atoms and molecules}

\date{\today}

\author{P\'eter Jeszenszki} 
\author{D\'avid Ferenc} 
\author{Edit M\'atyus} 
\email{edit.matyus@ttk.elte.hu}
\affiliation{ELTE, Eötvös Loránd University, Institute of Chemistry, 
Pázmány Péter sétány 1/A, Budapest, H-1117, Hungary}

\begin{abstract}
\noindent %
The Dirac--Coulomb equation with
positive-energy projection is solved using explicitly correlated Gaussian functions. The algorithm and computational procedure aims for a parts-per-billion convergence of the energy to provide a starting point for further comparison and further developments in relation with high-resolution atomic and molecular spectroscopy.
Besides a detailed discussion of the implementation of the fundamental spinor structure,  permutation and point-group symmetries, various options for the positive-energy projection procedure are presented. 
The no-pair Dirac--Coulomb energy converged to a parts-per-billion precision is compared with perturbative results for atomic and molecular systems with small nuclear charge numbers.
The subsequent paper [Paper~II: D. Ferenc, P. Jeszenszki, and E. Mátyus (2022)] describes the implementation of the Breit interaction in this framework.
\end{abstract}
\maketitle

\section{Introduction}
\noindent % 
For a quantitative description of the high-resolution spectroscopic measurements of atoms \cite{matveev_precision_2013,gurung_precision_2020} and molecules \cite{holsch_benchmarking_2019,semeria_precision_2020}, calculations of highly accurate energies are required corresponding to an at least parts-per-billion (ppb) relative precision \cite{puchalskiCompleteAlpha6m2016,puchalskiNonadiabaticQEDCorrection2019a,ferencNonadiabaticRelativisticLeadingOrder2020}. 
To ensure the ppb level of convergence for atomic and molecular energies, it is necessary to use explicitly correlated basis functions  \cite{mitroyTheoryApplicationExplicitly2013,puchalskiNonadiabaticQEDCorrection2019a,matyusMolecularStructureCalculations2012a,ferencNonadiabaticMassCorrection2019}. 
Although fast convergence of the energy with respect to the basis set size is ensured, most explicitly correlated functions are highly specialised to the particular system. The integral expressions for the matrix elements typically depend on the number of electrons and nuclei, and new derivations are required for every extra atom and molecule type \cite{frommAnalyticEvaluationThreeelectron1987,pachuckiCorrelatedExponentialFunctions2012,wangRelativisticCorrectionsGround2018}. 

Variants of explicitly correlated Gaussian functions (ECGs) \cite{jeziorskiHighaccuracyComptonProfile1979,suzukiStochasticVariationalApproach1998,mitroyTheoryApplicationExplicitly2013} have the advantage that they explicitly contain the interparticle distances, while they preserve the general analytic formulation for a variety of systems \cite{suzukiStochasticVariationalApproach1998,stankeAlgorithmsCalculatingMassvelocity2016,stankeOrbitorbitRelativisticCorrection2016}. 
In spite of these favourable properties, the ECGs are smooth functions that fail to satisfy the exact particle coalescence properties (cusps) of the non-relativistic wave function \cite{katoEigenfunctionsManyparticleSystems1957a} and the cuspy or singular coalescence points corresponding to relativistic model Hamiltonians, for which the precise properties depend on the particle interaction type \cite{kutzelniggGeneralizationKatoCusp1989,liRelativisticExplicitCorrelation2012}.
For certain non-relativistic quantities, there exists convergence acceleration techniques that account for the missing cusp effects \cite{pachuckiAccelerationConvergenceSingular2005,jeszenszkiInclusionCuspEffects2021}, while for variational relativistic treatments careful convergence tests and comparison with other specialized methods \cite{pestkaComplexCoordinateRotation2007,bylickiRelativisticHylleraasConfigurationinteraction2008}, which account for the singular behaviour, are relevant (see also Sec.~\ref{sec:heatom} of this work and Sec.~III~B of~Paper~II \cite{ferencBreitInteractionExplicitly2021}).

Regarding the physical model, for systems with small $Z$ nuclear charge numbers, the non-relativistic QED (nrQED) approach is commonly used, which includes the $Z\alpha$ expansion of the Dirac Hamiltonian \cite{pachuckiNonrelativisticQEDApproach2004,pachuckiHigherorderEffectiveHamiltonian2005,puchalskiCompleteAlpha6m2016,patkosRadiativeBmAlpha2021} and $\alpha$ is the fine-structure constant.
In the nrQED procedure, the relativistic corrections appear in the perturbative terms without any direct account for the interaction of electron correlation with relativistic and QED effects at lowest order.
There is another limiting range for which a meaningful expansion can be carried out, the $1/Z$ expansion is a common choice for high $Z$ values relevant for heavy elements \cite{mohrQuantumElectrodynamicsHigh1985,shabaevTwotimeGreenFunction2002,volotkaManyElectronQEDCorrections2014,yerokhinElectroncorrelationEffectsFactor2017}.  
The `intermediate' range, with intermediate $Z$, is a challenging range for the theory, because in this range, both the correlation and the relativistic (QED) effects are important \cite{indelicatoQEDRelativisticCorrections2007,malyshevQEDCalculationGroundstate2014}. 

In the present work, we will consider an approach that aims for a treatment of electron correlation and special relativity on the `same footing', and at the same time, targets tight convergence for the computed energies. In this way, comparison with results of the nrQED methodology, which can be considered well established for the low-$Z$ range, becomes relevant. 
In particular, an nrQED calculation is always restricted to a given order of the perturbation theory, and finding a good estimate for the contributions from higher orders is often challenging.

The Dirac--Coulomb (DC) Hamiltonian is often cited as a starting point to account for special relativity in a many-particle atomic or molecular system. In this model, the direct product of one-particle Dirac operators and the Coulomb interaction between the particles is considered. 
Energies and wave functions for this seemingly \emph{ad hoc} construct may be obtained by diagonalization of the matrix representation of the Dirac--Coulomb operator. Unfortunately, this simple procedure is problematic: electronic states, which would represent bound states, are `dissolved' in the positron-electron continuum in the infinite basis limit. This problematic behaviour is called the `continuum dissolution' or `Brown--Ravenhall (BR) disease' \cite{brownInteractionTwoElectrons1951}. 
The usual strategy to have access to these states relies on a positive--energy projection of the operator \cite{josephsucherEnergyLevelsTwoelectron1958,sucherFoundationsRelativisticTheory1980}, which eliminates the positron-positron and positron-electron states. 
Although, the projected energies depend on the projector, and hence on the underlying non-interacting model \cite{heullyCommentRelativisticWave1986,mittlemanTheoryRelativisticEffects1981,almoukhalalatiElectronCorrelationRelativistic2016},  the final energies and predicted spectroscopic quantities should be independent of these technical details, when all relevant QED terms are also accounted for \cite{sucherFoundationsRelativisticTheory1980,sucherFoundationsRelativisticTheory1983,dyallIntroductionRelativisticQuantum2007}. 

Using a determinant expansion, several approximate \cite{indelicatoQEDRelativisticCorrections2007,shabaevQEDMODFortranProgram2015} and {\it ab initio} \cite{malyshevQEDCalculationGroundstate2014,volotkaQEDRadiativeCorrections2019} QED computations have been carried out  using the Dirac--Coulomb(--Breit) model used as the zeroth-order Hamiltonian. 
At the same time, far fewer applications have been reported with explicitly correlated basis functions, partly  due to the difficulty caused by the missing one-particle picture in this basis representation.
A one-particle basis provides a common and natural starting point for the description of the interaction of elementary particles  \cite{l.n.labzokskiiElectronCorrealtionRelativistic1971,shabaevDualKineticBalance2004,lindgrenRelativisticManyBodyTheory2011,indelicatoIntroductionBoundStateQuantum2017}, and it can be used to develop a second quantized framework emphasized by Liu and co-workers \cite{liuPerspectivesRelativisticQuantum2012,liuGoingNopairRelativistic2013,liuAdvancesRelativisticMolecular2014,shaoBasicStructuresRelativistic2017}. 
At the same time, `explicit correlation' is required to accurately describe particle correlation that is important for spectroscopic applications \cite{cencekAccurateRelativisticEnergies1996,puchalskiCompleteAlpha6m2016,patkosRadiativeBmAlpha2021}. Liu and co-workers have considered the explicit correlation as a perturbation in the F12 framework, which required a `dual-basis' generalization of their positive-energy projection approach \cite{liRelativisticExplicitCorrelation2012,liuPerspectivesRelativisticQuantum2012,liuRelativisticExplicitCorrelation2017}.

Apart from our earlier report \cite{jeszenszkiAllorderExplicitlyCorrelated2021}, the very first and so far, single, implementation of positive-energy projection with an explicitly correlated basis set
was based on complex scaling of the particle coordinates \cite{bylickiRelativisticHylleraasConfigurationinteraction2008}. The complex scaling of the non-interacting model results in electronic states that are rotated 
to a branch in the complex-energy plane that is separated from the electron-positron and positron-positron states. In this way, the (non-interacting) electronic states can be expressed in a non-separable basis, and the corresponding  positive-energy functions can be identified. 
Bylicki, Pestka, and Karwowski \cite{bylickiRelativisticHylleraasConfigurationinteraction2008} proposed and implemented this complex-scaling approach with a(n explicitly correlated) Hylleraas basis set and used it to compute the ground-state Dirac--Coulomb energy of the isoelectronic series of the helium atom.

In the present work, we report the detailed theoretical background for the first extension of an explicitly correlated, positive-energy projected approach to molecules \cite{jeszenszkiAllorderExplicitlyCorrelated2021}.
We report the theoretical and algorithmic details for the computation of energies and eigenfunctions of the positive-energy projected or, as it is also called, no-pair DC Hamiltonian. After introduction of the DC model, the methodology is presented for the ECG framework, with explanation about the implementation of the permutational and point-group symmetries. We discuss in detail the positive-energy projection approach, for which three alternatives are considered in detail. 
The paper ends with the presentation and analysis of the numerical results in comparison with perturbative relativistic and QED energies for the example of low-$Z$ atomic and molecular systems.

\section{Dirac--Coulomb Hamiltonian and Dirac-spinor for two particles}
\subsection{Dirac--Coulomb Hamiltonian}
To introduce notation, we first consider the eigenvalue equation for the Dirac Hamiltonian (written in Hartree atomic units) of a single electron in interaction with $\nnuc$ fixed nuclei described as positive point charges,
\begin{align}
     h_\mathrm{D}^{[4]} \,  \varphi^{(4)} &= E \, \varphi^{(4)} \ , \label{Diraceq} \\
    h_\mathrm{D}^{[4]} &= c \balpha^{[4]} \bos{p} + \beta^{[4]} m c^2 + 1^{[4]} U \ ,  \label{Diracop} \\
    U &= - \sum_{I=1}^{\nnuc}  \frac{Z_I}{\left|\bos{r}-\bos{R}_I \right| }\ .
\    
    \label{Potop}
\end{align}
In the equations, $\bos{r}$ is the position of the electron, $\bos{R}_I$ is the position of the $I$-th nucleus, $Z_I$ is the charge of this nucleus, $1^{[n]}$ is the $n$-dimensional unit matrix, $\varphi^{(4)}$ is a four-component spinor, $\balpha^{[4]}=(\alpha_1^{[4]},\alpha_2^{[4]},\alpha_3^{[4]})$. Throughout this work, we use the superscript `$(n)$' and `$[n]$' to label an $n$-dimensional vector and an $(n\times n)$-dimensional matrix, respectively. The Dirac matrices are chosen according to the usual convention,
\begin{align}
    \alpha_i^{[4]} = \left( \begin{array}{cc}
    0^{[2]}  &  \sigma_i^{[2]} \\
    \sigma_i^{[2]} & 0^{[2]}
    \end{array}\right) \ , \hspace{2cm} 
    \beta^{[4]} = \left( \begin{array}{cc}
    1^{[2]} & 0^{[2]} \\
    0^{[2]} & -1^{[2]}
    \end{array}\right) \  \label{alphabeta}
\end{align}
with the $\sigma_i^{[2]}$ Pauli matrices,
\begin{align}
  \sigma_1^{[2]}
  =
  \left(%
    \begin{array}{@{}cc@{}}
      0 & 1 \\
      1 & 0 \\
    \end{array}
  \right)\ ,
  \; 
  \sigma_2^{[2]}
  =
  \left(%
    \begin{array}{@{}cc@{}}
      0 & -\iim \\
      \iim & 0 \\
    \end{array}
  \right)\ ,
  \ \text{and} \
  \; 
  \sigma_3^{[2]}
  =
  \left(%
    \begin{array}{@{}cc@{}}
      1 & 0 \\
      0 & -1 \\
    \end{array}
  \right)\ .
\end{align}
The matrix $0^{[n]}$ is the $(n\times n)$-dimensional zero matrix. 
According to the $2\times 2$ block structure of the $\alpha^{[4]}$ and $\beta^{[4]}$ matrices, Eq.~(\ref{alphabeta}), it is convenient to express the $\varphi^{(4)}$ spinor with a $\varphi^{\mathrm{l}(2)}$ `large' and a $\varphi^{\mathrm{s}(2)}$ `small' component, as
\begin{align}
    \varphi^{(4)}(\bos{r}) = \left[ \begin{array}{c}
    \varphi^{\mathrm{l}(2)} (\bos{r}) \\
    \varphi^{\mathrm{s}(2)} (\bos{r})
    \end{array}
    \right] \ . 
\end{align}
Both $\varphi^{\mathrm{l}(2)}$ and $\varphi^{\mathrm{s}(2)}$ have two components that can be characterized according to the spin projection on the $z$ axis ($+1/2$: $\uparrow$ and $-1/2$: $\downarrow$),
\begin{align}
    \varphi^{\lambda(2)} (\bos{r}) = \left[ \begin{array}{c}
    \varphi^{\lambda}_{\uparrow} (\bos{r}) \\
    \varphi^{\lambda}_{\downarrow} (\bos{r})
    \end{array}
    \right] \; ,\quad \lambda=\text{l or s} \; .
\end{align}
The (exact) relation of the large and the small components is obtained from \rrefsa{Diraceq}--\rrefsb{alphabeta},
\begin{align}
    \varphi^{\mathrm{s}(2)}(\bos{r}) = \frac{c \bos{\sigma}^{[2]} \bos{p}}{E-U+mc^2}  \varphi^{\mathrm{l}(2)}(\bos{r}) \ \;.\label{atombal}
\end{align}
To compute low-lying positive energy states that appear a bit below $mc^2$, this relation is commonly approximated by using $E-U+mc^2\approx 2mc^2$,
%$mc^2\gg U$ and $E\approx mc^2$,
\begin{align}
    \varphi^{\mathrm{s}(2)}(\bos{r}) \approx  \frac{ \bos{\sigma}^{[2]} \bos{p}}{2 mc}  \varphi^{\mathrm{l}(2)}(\bos{r}) \label{kinbal}\ .
\end{align}
It is important to note the symmetry relation (opposite parities) between the large and the small components, which is nicely discussed in Ref.~\cite{liuIdeasRelativisticQuantum2010}. 

We use this, so-called `restricted', kinetic balance condition \cite{kutzelniggBasisSetExpansion1984,dyallExactSeparationSpin1994,liuIdeasRelativisticQuantum2010} in the sense of a metric, 
\begin{align}
    \varphi^{(4)}(\bos{r}) &= \sum_{i=1}^{N_b} \sum_{q=1}^{4}c_{iq} \phi_{iq}^{(4)}(\bos{r}) \ , \\
    \phi_{iq}^{(4)}(\bos{r}) &= \mathcal{B}^{[4]} \,  1^{(4)}_q \, \Theta_i(\bos{r})  \ , \label{onepartbas} \\ 
    \mathcal{B}^{[4]}&=\left( \begin{array}{cc} 
    1^{[2]} & 0^{[2]} \\
    0^{[2]} & \frac{\bos{\sigma}^{[2]} {p}}{2mc}
    \end{array}\right) \label{onepartB}\ ,
\end{align}
where $c_{iq}$ is a coefficient, $\Theta_i(\bos{r})$ is a spatial function, and $1^{(4)}_q$ is a four-dimensional vector, in which all elements are zero except for the $q$-th element that is one,  $\left(1^{(4)}_q\right)_i=\delta_{qi}$.
This relation has a central importance for the construction of a good matrix representation of the Hamiltonian in numerical computations, otherwise
a `variational collapse'  would occur caused by an inappropriate representation of the $\bos{p}$ momentum operator in the spinor basis \cite{schwarzTwoProblemsConnected1982,kutzelniggCompletenessKineticallyBalanced2007,kutzelniggBasisSetExpansion1984}.

We also note that the rigorous variational property of the physical ground state would be guaranteed, in a strict mathematical sense, only by the `atomic balance' 
\cite{lewinSpectralPollutionHow2010},  which reads as,% \rref{atombal},
\begin{align}
    \varphi^{\mathrm{s}(2)}(\bos{r}) \approx \frac{c \bos{\sigma}^{[2]} \bos{p}}{2mc^2-U}  \varphi^{\mathrm{l}(2)}(\bos{r}) \ \;.
\end{align}
Unfortunately, application of the atomic balance would result (with practical basis sets) in matrix elements that are difficult (impossible) to integrate analytically and already the `restricted' kinetic balance, Eq.~(\ref{kinbal}),% \jadd{, moreover, there is an uncertainty due to a possible gauge dependence [Ref]. The 'restricted' kinetic balance, Eq.~(\ref{kinbal}), is gauge independent and}  
provides excellent results. 
So, in the present work, we will proceed with the `restricted' kinetic balance condition, Eq.~(\ref{kinbal}), and keep in mind the formal mathematical results.

For many-particle systems, several types of kinetic balance conditions have been introduced, which have different advantages depending on the aim of the computation \cite{kutzelniggBasisSetExpansion1984,lewinSpectralPollutionHow2010,kutzelniggSolvedUnsolvedProblems2012,simmenRelativisticKineticbalanceCondition2015}.
Shabaev and co-workers defined the dual kinetic balance condition \cite{shabaevDualKineticBalance2004,sunComparisonRestrictedUnrestricted2011} that implements not only the large-small, but also the small-large relation. Pestka, Bylicki,and Karwowski \cite{pestkaHylleraasCIApproachDiraccoulomb2003,bylickiRelativisticHylleraasConfigurationinteraction2008} mention an iterative procedure connecting the large and small subspaces.
Reiher and co-workers introduced a many-particle, so-called `relativistic' kinetic balance condition \cite{simmenRelativisticKineticbalanceCondition2015} by solving the two(many)-electron equations by using the $E-U+mc^2\approx 2 mc^2$ approximation. 

In a many-electron (many-spin-1/2-fermion) system, the Dirac operator for the $i$th particle is written in a direct-product form
\begin{align}
  h_i^{[4^N]} =  \unitfour(1)\boxtimes \ldots\boxtimes h_\mathrm{D}^{[4]}(i) \otimes \ldots \boxtimes \unitfour(N) \ ,
  \label{eq:hmulti}
\end{align}
where the particle index is given in parenthesis  and $N$ 
is the total number of electrons. 
By assuming instantaneous Coulomb interactions acting between the pairs of particles, we can write down 
 the eigenvalue equation,
\begin{align}
    \mathcal{H}_\mathrm{DC}^{[4^N]} \left| \Psi^{(4^N)} \right \rangle &= E \left| \Psi^{(4^N)}  \right \rangle \ ,  \label{DCeigeneq} 
\end{align}
with the Dirac--Coulomb Hamiltonian, 
\begin{align}
    \mathcal{H}_\mathrm{DC}^{[4^N]} &= \sum_i^N h_i^{[4^N]} + 1^{[4^N]} V \ , \\[0.5cm]
     V &= 
     \sum_{i=1}^N \sum_{i<j}^N 
     \frac{1}{\left| \bos{r}_i-\bos{r}_j\right|}\ . 
\end{align}
In the many-particle case, it remains to be convenient to think in terms of the large-small block structure similarly to the one-electron case. The many-particle spinor has in total $4^N$ components that is now  considered in terms of $2^N$ large-small components and $2^N$ spin configurations. The block-wise direct product, which allows us to retain the large-small structure, was called the Tracy--Singh product \cite{tracyNewMatrixProduct1972} in Refs.~\cite{liRelativisticExplicitCorrelation2012,shaoBasicStructuresRelativistic2017} and was later also used in Refs.~\cite{simmenRelativisticKineticbalanceCondition2015,jeszenszkiAllorderExplicitlyCorrelated2021}. 

In this paper, we focus on  two-electron systems, for which the block-wise spinor structure can be written as
\begin{align}
\left|  \Psi^{(16)} \right \rangle  &= 
\begin{bmatrix}
\left| \psi^{\mathrm{l l}(4)} \right \rangle \\[0.2cm]
\left| \psi^{\mathrm{l s}(4)} \right \rangle \\[0.2cm]
\left| \psi^{\mathrm{s l}(4)} \right \rangle\\[0.2cm]
\left| \psi^{\mathrm{s s}(4)} \right \rangle 
\end{bmatrix} 
\end{align}
and
\begin{align}
  \left| \psi^{\lambda_1 \lambda_2 (4)} \right \rangle  &=
\begin{bmatrix}
\left| \psi^{\lambda_1 \lambda_2}_{\uparrow \uparrow} \right \rangle   \\[0.2cm]
\left| \psi^{\lambda_1 \lambda_2}_{\uparrow \downarrow} \right \rangle  \\[0.2cm]
\left| \psi^{\lambda_1 \lambda_2}_{\downarrow \uparrow} \right \rangle \\[0.2cm]
\left| \psi^{\lambda_1 \lambda_2}_{\downarrow \downarrow} \right \rangle   
\end{bmatrix} \ , 
\label{spinor}
\end{align}
where $\lambda_1$ and $\lambda_2$ can be `$\mathrm{l}$' or `$\mathrm{s}$'.
The two-particle Dirac--Coulomb Hamiltonian written in a corresponding block-structure is
\begin{widetext}

\begin{align}
     & \mathcal{H}_\mathrm{DC}^{[16]}  = \label{eq:HDC2p}\\
     &  \left(%
       \begin{array}{cccc}
           \left( V+U \right) 1^{[4]} & 
          c\bos{\sigma}_2^{[4]}\bos{p}_2  & 
          c\bos{\sigma}_1^{[4]} \bos{p}_1 & 
          0^{[4]}  \\
          c\bos{\sigma}_2^{[4]}\bos{p}_2  &  
          \left( V+U -2m_2 c^2\right) 1^{[4]}   & 
          0^{[4]} & 
         c\bos{\sigma}_1^{[4]} \bos{p}_1\\
          c\bos{\sigma}_1^{[4]} \bos{p}_1  &  
          0^{[4]} & 
          \left( V+U - 2m_1 c^2 \right) 1^{[4]}   & 
          c\bos{\sigma}_2^{[4]}\bos{p}_2   \\
          0^{[4]}  & 
          c\bos{\sigma}_1^{[4]} \bos{p}_1  &  
          c\bos{\sigma}_2^{[4]}\bos{p}_2   & 
          \left[ V+U  - 2\left(m_1+ m_2\right)c^2  \right] 1^{[4]}
       \end{array}
     \right) \ ,  \nonumber
\end{align}
\end{widetext}
where, similarly to \rref{eq:hmulti}, the
$\bos{\sigma}_1^{[4]}=\bos{\sigma}^{[2]}\otimes 1^{[2]}$ notation is introduced
and 
$\bos{\sigma}_2^{[4]}=1^{[2]}\otimes \bos{\sigma}^{[2]}$ and the energy scale for the $i$th particle is shifted by $m_i c^2$.
We use during this work the relationship 
\begin{align}
  \bos{\sigma}^{[4]}_1\bos{\sigma}^{[4]}_2
  =
  (\bos{\sigma}^{[2]}\otimes 1^{[2]})
  (1^{[2]}\otimes \bos{\sigma}^{[2]})
  =
  \bos{\sigma}^{[2]} \otimes \bos{\sigma}^{[2]} \; .
\end{align}
The exact wave function is expanded in a spinor basis, 
\begin{align}
 | \Psi^{(16)} \rangle = \sum_{i=1}^{N_\text{b}} \sum_{q=1}^{16} c_{iq}  | \Phi_{iq}^{(16)} \rangle  \label{basisetexp}  \; ,
\end{align}
for which the kinetic-balance condition of \rref{onepartbas} can be generalized \cite{kutzelniggBasisSetExpansion1984,dyallExactSeparationSpin1994} as
\begin{align}
    | \Phi_{iq}^{(16)} \rangle& = \mathcal{B}^{[16]} \, 1_q^{(16)}  \, | \Theta_i \rangle \ , \label{spinoransatz} \\
    \mathcal{B}^{[16]}& = \left( \begin{array}{cccc} 
    1^{[4]} & 0^{[4]} & 0^{[4]} & 0^{[4]} \\
    0^{[4]} & \frac{\bos{\sigma}^{[4]}_2 \bos{p}_2}{2m_2c} & 0^{[4]} & 0^{[4]} \\
    0^{[4]} & 0^{[4]} & \frac{\bos{\sigma}^{[4]}_1 \bos{p}_1}{2mc_1}  & 0^{[4]} \\
    0^{[4]} & 0^{[4]} & 0^{[4]} & \frac{ \bos{\sigma}^{[4]}_1 \bos{p}_1  \bos{\sigma}^{[4]}_2 \bos{p}_2}{4m_1m_2c^2}  
    \end{array}\right) \; .
    \label{eq:kineticBalance}
\end{align}
Additionally, $| \Theta_i \rangle$ is a floating explicitly correlated Gaussian function, 
\begin{align}
        \Theta_i(\br) &= \exp \left[-\left( \br-\bs_i \right)^T \underline{\bA}_i \left( \br - \bs_i \right)\right] \ ,
    \label{eq:ECGansatz}
\end{align}
where $\br\in\mathbb{R}^6$ is the position vector of the two particles, $\bs_i\in\mathbb{R}^6$ is the `shift' vector, and $\underline{\bA}_i=\bA_i\otimes 1^{[3]}$ with $\bA_i\in\mathbb{R}^{2\times 2}$ is a symmetric, positive-definite parameter matrix.

Multiplying \rref{DCeigeneq} from the left with $\left\langle \Phi_{jp}^{(16)} \right|$ and using \rrefsa{basisetexp} and \rrefsb{spinoransatz}, we obtain the matrix eigenvalue equation, 
\begin{align}
    &\sum_{i=1}^{N_b} \sum_{q=1}^{16}  \left \langle \Phi_{jp}^{(16)} \left| \mathcal{H}_\mathrm{DC}^{[16]} \right | \Phi_{iq}^{(16)} \right \rangle c_{iq}  = %\nonumber \\ 
    %& \hspace{3cm}
    E  \sum_{i=1}^{N_b} \sum_{q=1}^{16} \left \langle \Phi_{jp}^{(16)} \right. \left| \Phi_{iq}^{(16)} \right \rangle c_{iq} 
    \ .
    \label{eq:mxeigenalue}
\end{align}
The explicit form for a matrix element is 
\begin{widetext}
\begin{align}
     &\langle \Phi_{jp}^{(16)} | \mathcal{H}_\mathrm{DC}^{[16]} | \Phi_{iq}^{(16)} \rangle =  \langle \Theta_j | {1^{(16)}_p}^T {\mathcal{B}^{[16]}}^\dagger \mathcal{H}_\mathrm{DC}^{[16]} \mathcal{B}^{[16]} 1^{(16)}_q  | \Theta_i \rangle = \label{eq:dcmx}\\
     & {1^{(16)}_p}^T  \left(%
       \begin{array}{cccc}
          \left \langle \Theta_j \left| \mathcal{W}_\mathrm{ll}^{[4]}  \right| \Theta_i \right \rangle & 
          \frac{1}{2m_2}\langle \Theta_j \left| p_2^2 \right | \Theta_i \rangle 1^{[4]}  & 
          \frac{1}{2m_1}\langle \Theta_j \left| p_1^2  \right | \Theta_i \rangle 1^{[4]} & 
          0^{[4]}  \\
          \frac{1}{2m_2}\langle \Theta_j \left| p_2^2 \right | \Theta_i \rangle 1^{[4]} &  
           \left \langle \Theta_j \left| \mathcal{W}_\mathrm{ls}^{[4]}  \right| \Theta_i \right \rangle & 
          0^{[4]} & 
          \frac{1}{8m_1m_2^2c^2}\langle \Theta_j \left| p_1^2 p_2^2 \right | \Theta_i \rangle 1^{[4]} \\
          \frac{1}{2m_1} \langle \Theta_i \left| p_1^2 \right | \Theta_j \rangle 1^{[4]}  &  
          0^{[4]} & 
          \left \langle \Theta_j \left| \mathcal{W}_\mathrm{sl}^{[4]}  \right| \Theta_i \right \rangle  & 
          \frac{1}{8m_1^2m_2c^2} \langle \Theta_i \left| p_1^2 p_2^2 \right | \Theta_j \rangle 1^{[4]}  \\
          0^{[4]}  & 
          \frac{1}{8m_1m_2^2c^2}\langle \Theta_i \left| p_1^2 p_2^2 \right | \Theta_j \rangle 1^{[4]}  &  
          \frac{1}{8m_1^2m_2c^2}\langle \Theta_i \left|p_1^2  p_2^2 \right | \Theta_j \rangle 1^{[4]}  & 
           \left \langle \Theta_j \left| \mathcal{W}_\mathrm{ss}^{[4]}  \right| \Theta_i \right \rangle
       \end{array}
     \right) 1^{(16)}_q \ ,  \nonumber
     \end{align}
     \begin{align}
     &\langle \Phi_{jp}^{(16)}  | \Phi_{iq}^{(16)} \rangle = \langle \Theta_j | {1^{(16)}_p}^T  {\mathcal{B}^{[16]}}^\dagger \mathcal{B}^{[16]} 1^{(16)}_q  | \Theta_i \rangle =  \label{eq:overmx}  \\
     & {1^{(16)}_p}^T  \left(
     \begin{array}{cccc}
          \langle \Theta_i  | \Theta_j \rangle 1^{[4]} & 0^{[4]} & 0^{[4]} & 0^{[4]}  \\
           0^{[4]} &  \frac{1}{4m_2^2c^2} \langle \Theta_i  \left| p_2^2  \right | \Theta_j \rangle 1^{[4]} & 0^{[4]} & 0^{[4]} \\
            0^{[4]} & 0^{[4]} &  \frac{1}{4m_1^2c^2} \langle \Theta_i  \left| p_1^2 \right | \Theta_j \rangle 1^{[4]}  & 0^{[4]} \\
          0^{[4]} & 0^{[4]} & 0^{[4]} &  \frac{1}{16m_1^2m_2^2c^4} \langle \Theta_i  \left| p_1^2 p_2^2 \right | \Theta_j \rangle  1^{[4]} 
     \end{array}
     \right) 1^{(16)}_q \ ,
     %\label{eq:phiiphij}
     \nonumber\\  
       \mathcal{W}_\mathrm{ll}^{[4]}=&  \left( V+ U \right) 1^{[4]}  \ ,  \\
      \mathcal{W}_\mathrm{ls}^{[4]}=&\frac{1}{4m_2^2c^2} \sum_{i,j=1}^3   p_{2i} \left( V+U\right) p_{2j}  \sigma_{2i}^{[4]} \sigma_{2j}^{[4]}  -  \frac{1}{2m_2} p_2^2  1^{[4]}  \ ,  \\
      \mathcal{W}_\mathrm{sl}^{[4]}=&\frac{1}{4m_1^2c^2} \sum_{i,j=1}^3 p_{1i} \left( V+ U\right) p_{1j} \sigma_{1i}^{[4]}  \sigma_{1j}^{[4]}  -  \frac{1}{2m_1} p_1^2  1^{[4]}  \ ,  \\
     \mathcal{W}_\mathrm{ss}^{[4]}=&\frac{1}{16m_1^2m_2^2c^4}\sum_{i,j,k,l=1}^3   p_{1i} p_{2j}  \left( V + U \right) p_{2k} p_{1l} \sigma_{1i}^{[4]} \sigma_{1l}^{[4]}  \sigma_{2j}^{[4]} \sigma_{2k}^{[4]}   -\frac{(m_1+m_2)}{8m_1^2 m_2^2 c^2}   p_1^2 p_2^2  1^{[4]}  \ ,  \label{Wss}
\end{align}
\end{widetext}
where the $(\bos{\sigma}^{[4]}\bos{p})(\bos{\sigma}^{[4]}\bos{p}) = p^2 1^{[4]}$ identity was used.
In the equations, $\left \langle \Theta_i  \left| \mathcal{W}^{[4]} \right| \Theta_j \right \rangle$ means that the integral is evaluated for every element of the four-dimensional matrix.
The next subsection describes the implementation of the permutational symmetry for (two) identical spin-1/2 particles.

\subsection{Implementation of the permutational symmetry\label{sec:permutsym}}
The Dirac--Coulomb Hamiltonian, similarly to its non-relativistic counterpart, is invariant to the permutation of identical particles. At the same time, it is necessary to consider the multi-dimensional spinor structure when we calculate the effect of the particle-permutation operator. For two particles and for the present block structure, it is 
\begin{align}
  \mathcal{P}_{12}^{[16]} 
  = 
  \left(\mathbb{P}^{\mathrm{ls}[4]}_{12} 
  \otimes 
  \mathbb{P}^{\uparrow \downarrow [4]}_{12} \right) P_{12} \ , \label{pairPidef} 
\end{align}
where 
$P_{12}$ exchanges the particle labels (as in the non-relativistic theory), but due to the multi-dimensional spinor structure, we also have
$\mathbb{P}^{\mathrm{ls}[4]}_{12}$, which operates on the large-small component space, and $\mathbb{P}^{\uparrow \downarrow[4]}_{12}$, which acts on the $\uparrow-\downarrow$ spin space, and both of them are represented with the following matrix
\begin{align}
    \mathbb{P}^{\text{ls} [4]}_{12}
    =
    \mathbb{P}^{\uparrow \downarrow [4]}_{12}
    =
    \left(%
    \begin{array}{cccc}
        1 & 0 & 0 & 0 \\
        0 & 0 & 1 & 0 \\
        0 & 1 & 0 & 0 \\
        0 & 0 & 0 & 1
    \end{array}\right) \ .
\end{align}
The permutation operator, Eq.~(\ref{pairPidef}), 
commutes with the Dirac--Coulomb Hamiltonian  \cite{shaoBasicStructuresRelativistic2017,dyallIntroductionRelativisticQuantum2007},
\begin{align}
  [ \mathcal{H}_\DC^{[16]},\mathcal{P}_{12}^{[16]}]=0 \; ,
\end{align}
and thus, their common eigenfunctions can be identified as bosonic (symmetric) or fermionic (antisymmetric) eigenstates.
Since we describe spin-1/2 fermions, it is convenient to work with antisymmetrized basis functions, for which we introduce the antisymmetrization operator \cite{dyallIntroductionRelativisticQuantum2007} that reads for the two-particle case as,
\begin{align}
    \mathcal{A}^{[16]}=\frac{1}{2} \left( 1^{[16]} - \mathcal{P}_{12}^{[16]} \right) \ . \label{twoantism}
\end{align}

To define a variational procedure, we combine the antisymmetrized wave function ansatz, the finite basis expansion in \rref{basisetexp}, and the eigenvalue equation in \rref{DCeigeneq}. Then, by multiplying the resulting operator eigenvalue equation with $\langle \Phi_{jp} |\mathcal{A}^{[16] \dagger}$ using \rref{spinoransatz}, we obtain the matrix-eigenvalue equation, 
\begin{align}
    \sum_{i=1}^{N_\text{b}} \sum_{q=1}^{16} 
    & \left \langle \Phi_{jp}^{(16)} \left| \mathcal{A}^{[16]\dagger} \mathcal{H}_\mathrm{DC} \mathcal{A}^{[16]} \right| \Phi_{iq}^{(16)} \right \rangle c_{iq}  = 
    E 
      \sum_{i=1}^{N_\text{b}} \sum_{q=1}^{16}
    \left \langle \Phi_{jp}^{(16)} \left | \mathcal{A}^{[16]\dagger}  \mathcal{A}^{[16]} \right | \Phi_i^{(16)} \right \rangle c_{iq}
    \ . 
    \label{eq:eivaleq}
\end{align}
Since $\mathcal{A}^{[16]\dagger}=\mathcal{A}^{[16]}$, $\mathcal{A}^{[16]}$ is idempotent, and both $1^{[16]}$ and $\mathcal{P}_{12}^{[16]}$ commute with the Hamiltonian operator, \rref{eq:eivaleq} can be simplified to 
\begin{align}
    \sum_{i=1}^{N_\text{b}} \sum_{q=1}^{16} 
    & \left \langle \Phi_{jp}^{(16)} \left| \mathcal{H}_\mathrm{DC} \mathcal{A}^{[16]} \right| \Phi_{iq}^{(16)} \right \rangle c_{iq}  = 
    E 
      \sum_{i=1}^{N_\text{b}} \sum_{q=1}^{16}
    \left \langle \Phi_{jp}^{(16)} \left |  \mathcal{A}^{[16]} \right | \Phi_i^{(16)} \right \rangle c_{iq}
    \ . 
    \label{antismeqperm}
\end{align}

The permutation operator, $P_{12}$, acts only on the $\br$ spatial coordinates in $\Theta(\br)$ by exchanging the position of the particles. For a floating ECG, the effect of a particle permutation operator $P_{12}$ translates to 
a transformation of the ECG parameterization, \rref{eq:ECGansatz}, \cite{matyusMolecularStructureCalculations2012a,suzukiStochasticVariationalApproach1998}:
\begin{align}
  P_{12} \Theta(\br)
  &=
  \Theta(P^{-1}_{12} \br) \nonumber \\
  &=
  \exp \left[% 
    \left(P^{-1}_{12} \br-\bs_i \right)^\tT 
    \underline{\bA}_i 
    \left(P^{-1}_{12} \br - \bs_i \right)
  \right] \nonumber \\
  &=
  \exp\left[% 
    \left(\br- \underbrace{P_{12} \bs_i}_{\bs_i^{(12)}} \right)^\tT 
    \underbrace{P^\tT_{12} \underline{\bA}_i P_{12}}_{\underline{\bA}_i^{(12)}} 
    \left(\br-\underbrace{P_{12} \bs_i}_{\bs_i^{(12)}} \right)
  \right] \nonumber \\
  &=
  \exp\left[% 
    \left(\br-{\bs}_{i}^{(12)} \right)^\tT 
    \underline{{\bA}}_{i}^{(12)}
    \left(\br-{\bs}_{i}^{(12)} \right)
  \right] \; .
  \label{eq:permECGtransform}
\end{align}

\subsection{Implementation of the point-group symmetry \label{sec:spatsym}}
To have an efficient computational approach for molecules with clamped nuclei, it was necessary to make use of the point-group symmetry during the computations.
Due to the spinor structure of the Dirac equation, it is necessary to consider the spatial symmetry operations over the spinor space. Each symmetry operation, $\mathcal{S}^{[16]}$, will be considered as a composition of a symmetry operator $S$ acting on the configuration space and an operator
$\mathbb{S}^{[16]}$ acting on the spin part \cite{dyallIntroductionRelativisticQuantum2007},
\begin{align}
  \mathcal{O}^{[16]}= \mathbb{O}^{[16]} O \ .
\end{align}
In relativistic quantum mechanics, the notation of the spatial symmetry operators is often retained with a modified meaning that includes both spin and spatial symmetry operations. In the present work, we show both the spin and spatial contributions for clarity.
The spatial identity, rotation, and inversion symmetry operators for a single particle are generalized with the following operations for the spin-spatial space as \cite{dyallIntroductionRelativisticQuantum2007} 
\begin{align}
    \mathcal{E}^{[4]} &= 1^{[4]} E \ , \\
    \mathcal{C}_{na}^{[4]} &= 1^{[2]} \otimes \exp \left[- \iim \frac{\sigma_a^{[2]} \pi}{n} \right]  C_{na} \ , \\
    \mathcal{I}^{[4]} &= \beta^{[4]} I \ ,
\end{align}
where $a$ labels the  spatial directions. 
Further symmetry operations can be generated using these three symmetry operations. For example, the plane reflection can be obtained as an inversion followed by a two-fold rotation,
\begin{align}
    \mathcal{T}_{ab}^{[4]}= \mathcal{C}_{2c}^{[4]}  \mathcal{I}^{[4]}  \ , \label{reflection}
\end{align}
where $a$, $b$, and $c$ label mutually orthogonal spatial directions. Using Euler's formula for the spin part,
\begin{align}
  \exp \left[- \iim \frac{\sigma_a^{[2]} \pi}{n} \right]  = 1^{[2]} \cos \frac{\pi}{n} - \iim \sigma_a^{[2]} \sin \frac{\pi}{n} \ \label{spintrig} ,  
\end{align}
\rref{reflection} can be further simplified to
\begin{align}
    \mathcal{T}_{ab}^{[4]}= -\iim \left( 1^{[2]} \otimes \sigma_c^{[2]}   \mathcal{C}_{2c} \right) \left(\beta^{[4]} I \right) \ .
\end{align}

Since we solve the eigenvalue equation in the kinetic-balance metric defined by \rref{eq:kineticBalance}, it is more convenient to introduce 
an $\mathcal{O}_\mathcal{B}^{[4]}$ `modified' representation of the symmetry operators, 
\begin{align}
    \mathcal{O}_\mathcal{B}^{[4]} = \left(\mathcal{B}^{[4]}\right)^{-1} \mathcal{O} \mathcal{B}^{[4]} \ . \label{OBdef}
\end{align}
It can be shown that both $\mathcal{E}^{[4]}$ and $\mathcal{C}_{na}^{[4]}$ commutes with $\mathcal{B}^{[4]}$, hence, 
\begin{align}
    \mathcal{E}_\mathcal{B}^{[4]} &= \mathcal{E}^{[4]} \ , \\
    \mathcal{C}_{\mathcal{B},na}^{[4]} &= \mathcal{C}_{na}^{[4]} \ , 
\end{align}
but the inversion and reflection operators simplify to 
\begin{align}
    \mathcal{I_B}^{[4]} &= 1^{[4]} I \ , \\
     \mathcal{T}_{\mathcal{B},ab}^{[4]} &= -\iim \left( 1^{[2]} \otimes \sigma_c^{[2]}   \mathcal{C}_{2c} \right)  \mathcal{I} \ . \label{eq:modrefsym} 
\end{align}
For two particles, we need to consider the block-wise (Tracy--Singh) direct products,
\begin{align}
\mathcal{E}_\mathcal{B}^{[16]}&=1^{[16]} \mathcal{E}_1 \mathcal{E}_2 \ , \\
\mathcal{C}_{\mathcal{B},np}^{[16]} &= \left(  \eem^{-\iim \frac{\sigma_{p1}^{[2]} \pi }{n}} \otimes 1^{[2]} \right) \boxtimes  \left( 1^{[2]}  \otimes \eem^{-\iim \frac{\sigma_{p2}^{[2]} \pi}{n}} \right)  \mathcal{C}_{np,1} \mathcal{C}_{np,2} \ , \\
\mathcal{I}_\mathcal{B}^{[16]}  &= 1^{[16]} \mathcal{I}_1 \mathcal{I}_2 \ , \\
\mathcal{T}_{\mathcal{B},ab}^{[16]} &= -\left( \sigma_{c1}^{[2]}  \otimes 1^{[2]} \right) \boxtimes  \left( 1^{[2]}  \otimes \sigma_{c2}^{[2]}  \right)  \mathcal{C}_{2c,1} \,  \mathcal{C}_{2c,2} \, \mathcal{I}_1 \, \mathcal{I}_2 \ , \label{reflect2part}
\end{align}
where the subscripts 1 and 2 stand for the particle indices.

A (floating) ECG, \rref{eq:ECGansatz}, is adapted to the $\zeta$ irreducible representation of the point group $G$ by using the projector
\begin{align}
    P_G^{\zeta[16]} = \sum_{\mathcal{O} \in G} \chi_{G\mathcal{O}}^\zeta \mathcal{O}^{[16]} \ ,  \label{eq:projsym} 
\end{align}
where $\chi_{G\mathcal{O}}^\zeta$ labels the character corresponding to the $\mathcal{O}^{[16]}$ symmetry operation. In relativistic quantum mechanics, the point groups have to be extended to a double group for an odd number of particles. For the case of two half-spin particles, the simpler, well-known single-group character tables can be used \cite{saueQuaternionSymmetryRelativistic1999,dyallIntroductionRelativisticQuantum2007}.  

Due to the $\underline{\bA}_i=\bA_i\otimes I^{[3]}$ direct-product (`spherical-like') form of the exponent matrix, 
the effect of the spatial symmetry operators on the ECG function can be translated to the transformation of the shift vectors (see for example, Refs.~\cite{matyusMolecularStructureCalculations2012a,matyusPreBornOppenheimerMolecular2019}), 
\begin{align}
  \mathcal{O} \Theta(\br)
  &=
  \mathcal{O}
  \exp\left[% 
    \left(\br- \bs_i\right)^\tT 
    \underline{\bA}_i
    \left(\br-\bs_i\right)
  \right] 
  \nonumber \\
  &=
  \exp\left[% 
    \left(\br- \underbrace{\mathcal{O} \bs_i}_{\bs_i^{\mathcal{O}}} \right)^\tT 
    \underbrace{\mathcal{O}^\tT \underline{\bA}_i \mathcal{O}}_{\underline{\bA}_i} 
    \left(\br-\underbrace{\mathcal{O} \bs_i}_{\bs_i^{\mathcal{O}}} \right)
  \right]  
  \nonumber \\
  &=
  \exp\left[% 
    \left(\br- \bs_i^{\mathcal{O}} \right)^\tT
    \underline{\bA}_i
    \left(\br-\bs_i^{\mathcal{O}} \right)
  \right] 
  \label{eq:spatialECGtransform}
\end{align}
that allows a straightforward evaluation of the effect of the spatial symmetry operators.

In summary, the basis functions used in the computations are obtained by projection of the ECGs with the $\mathcal{A}^{[16]} \left| \Phi_i \right \rangle$ antisymmetrization and the $P_G^{\zeta[16]}$ point-group symmetry projector,  which results in the following generalized eigenvalue equation,
\begin{align}
        \sum_{i=1}^{N_\text{b}} \sum_{p=1}^{16}  &\left \langle \Phi_{jq}^{(16)} \left| \mathcal{H}_\mathrm{DC}^{[16]} P_G^{\zeta[16]} \mathcal{A}^{[16]} \right| \Phi_{ip}^{(16)} \right \rangle c_{ip}  = \label{eq:antismeigeq} \\
         &\hspace{1cm} 
         E  \sum_{i=1}^{N_\text{b}} \sum_{p=1}^{16} \left \langle \Phi_{jq}^{(16)} \left | P_G^{\zeta[16]} \mathcal{A}^{[16]} \right | \Phi_{ip}^{(16)} \right \rangle c_{ip}
    \ . \nonumber
\end{align}

Substituting  \rrefsa{spinoransatz}, \rrefsb{twoantism}, and \rrefsb{eq:projsym} into \rref{eq:antismeigeq} and using \rref{OBdef} and the fact that $\mathcal{B}^{[16]}$ and 
$\left(\mathbb{P}^{\mathrm{ls}[4]}_{12} \otimes \mathbb{P}^{\uparrow \downarrow [4]}_{12} \right) P_{12}$
commute, we obtain the final working equation as
\begin{widetext}
\begin{align}
    &\sum_{i=1}^{N_\text{b}} \sum_{p=1}^{16}
    \sum_{\mathcal{O} \in G} 
      \chi_{G\mathcal{O}}^\zeta \, {1_q^{(16)}}^T \Bigg[ \mathcal{K}_{ji}^{O[16]} \mathbb{O}_\mathcal{B}^{[16]}- \mathcal{K'}_{ji}^{O[16]}   
      \left(\mathbb{P}^{\mathrm{ls}[4]}_{12} \otimes \mathbb{P}^{\uparrow \downarrow [4]}_{12} \right)
      \mathbb{O}_\mathcal{B}^{[16]} \Bigg] 1_p^{(16)} c_{ip}  = \nonumber \\ 
    &\hspace{3.5cm} 
    E  
    \sum_{i=1}^{N_\text{b}} \sum_{p=1}^{16}
    \sum_{\mathcal{O} \in G} \chi_{G\mathcal{O}}^\zeta \, {1_q^{(16)}}^T \Bigg[   \mathcal{S}_{ji}^{O[16]} \mathbb{O}_B^{[16]} -  \mathcal{S'}_{ji}^{O[16]} 
    \left(\mathbb{P}^{\mathrm{ls}[4]}_{12} \otimes \mathbb{P}^{\uparrow \downarrow [4]}_{12} \right)
    \mathbb{O}_\mathcal{B}^{[16]} \Bigg] 1_p^{(16)} c_{ip} 
    \  \label{eq:workingeq}  \\
    & \mathcal{K}_{ji,qp}^{O[16]}= \left \langle \Theta_j \left|  {1_q^{(16)}}^T \mathcal{B}^{[16]\dagger} \mathcal{H}_\mathrm{DC}^{[16]} \mathcal{B}^{[16]} 1_p^{(16)} O\right| \Theta_i \right \rangle \label{Kmat}\\
    & \mathcal{K'}_{ji,qp}^{O[16]}= \left \langle \Theta_j \left|  {1_q^{(16)}}^T \mathcal{B}^{[16]\dagger} \mathcal{H}_\mathrm{DC}^{[16]} \mathcal{B}^{[16]} 1_p^{(16)} O P_{12}\right| \Theta_i \right \rangle \\
    & \mathcal{S}_{ji,qp}^{O[16]}=\left \langle \Theta_j \left|  {1_q^{(16)}}^T \mathcal{B}^{[16]\dagger} \mathcal{B}^{[16]} 1_p^{(16)} O\right| \Theta_i \right \rangle \\
    & \mathcal{S'}_{ji,qp}^{O[16]}=  \left \langle \Theta_j \left|  {1_q^{(16)}}^T \mathcal{B}^{[16]\dagger} \mathcal{B}^{[16]} 1_p^{(16)} O P_{12}\right| \Theta_i \right \rangle \; .\label{SmatP}
\end{align}
\end{widetext}
We note that in \rref{eq:workingeq} the dimensionality of the final matrix eigenvalue equation to be solved is $16N_\text{b} \times 16N_\text{b}$, since the linear combination coefficients are carried in
$c_{ip}\in\mathbb{C}$ $(i=1,\ldots,N_\text{b},p=1,\ldots,16)$.
For special cases, the matrix representation can be block diagonalized (a special case is described in the Supplementary Material).

\subsection{Implementation of the Dirac--Coulomb matrix-eigenvalue equation}

The working equation, \rref{eq:workingeq}, is implemented in the QUANTEN computer program
that is an in-house developed program written using the Fortran90 programming language and contains several analytic ECG integrals. For recent applications of  QUANTEN, see Refs.~\cite{matyusNonadiabaticMassCorrection2018,matyusNonadiabaticMasscorrectionFunctions2018,ferencComputationRovibronicResonances2019,ferencNonadiabaticMassCorrection2019,ferencNonadiabaticRelativisticLeadingOrder2020,matyusOrientationalDecoherenceMolecules2021,jeszenszkiInclusionCuspEffects2021,irelandLowerBoundsAtomic2021}. 
According to Eqs.~(\ref{eq:permECGtransform}) and (\ref{eq:spatialECGtransform}), the permutation and point-group projection operations can be translated to a parameter transformation of the ECG function, and do not require the calculation of (mathematically) new  spatial integrals. 
After construction of the matrices in \rrefsa{Kmat}--\rrefsb{SmatP}, the point-group symmetry projection is carried out by multiplication with the 16-dimensional (for two electrons) spinor part of the symmetry operator. The summation is performed over the symmetry elements ($\mathcal{O} \in G$) of the group according to \rref{eq:workingeq}.

Equation \rrefsb{eq:workingeq} contains both linear ($c_{ip}$, $i=1,\ldots,N_\text{b}$,  $p=1,\ldots,16$ ) and non-linear parameters ($\bA_i,s_i$, $i=1,\ldots,N_\text{b}$) in the basis functions, which can be  variationally optimized to improve the (projected or) no-pair relativistic energy. (We note also at this point that the currently used kinetic balance condition, \rref{kinbal}, is only an approximation to the mathematically rigorous `atomic' kinetic balance condition, Eq.~(\ref{atombal})  \cite{dolbeaultEigenvaluesOperatorsGaps2000,estebanVariationalMethodsRelativistic2008}.) The optimal linear parameters are obtained by solving $\bos{H}\bos{c}=E\bos{S}\bos{c}$-type generalized eigenvalue equation. 

To optimize the non-linear parameters, we minimize the energy (obtained by diagonalization) of the selected (ground or excited) physical state in the spectrum.
In contrast to the non-relativistic case, the physical ground-state energy is not the lowest-energy state of the `bare' (unprojected) DC Hamiltonian, lower-energy, often called `positronic' or `negative-energy' states also appear in the spectrum, hence, the electronic ground state is described by one of the excited states of the eigenvalue equation. Moreover, due to the Brown--Ravenhall disease \cite{brownInteractionTwoElectrons1951,karwowskiDiracOperatorIts2017}, 
the electronic ground state appears among the (finite-basis representation of the) electron-positron continuum states. 
In finite-basis computations and for small nuclear charge numbers, the electronic states are often found to be well separated from the positronic states and only a few positron-electron states `contaminate' the electronic spectrum. In this case, selection of the electronic states is possible based on a threshold energy that can be estimated by a value near (lower than) the non-relativistic energy. We list the computed `bare' DC energies in the Supplementary Material (Tables~S1--S4), but we consider them only as technical, computational details. The bare DC Hamiltonian is projected, and then, diagonalized to obtain the physically relevant results, \emph{e.g.,} the ground state energy as the lowest no-pair energy.

The optimization of the non-linear parameters by minimization of the energy for the selected DC state is a CPU-intensive part in the current implementation. Nevertheless, we ran several refinement cycles of the optimization procedure, but it hardly improved the DC energy in comparison with the DC results obtained with the non-linear (basis) parameters optimized by minimization of the corresponding non-relativistic energy (in low-$Z$ systems). At the same time, it is also worth noting that upon repeated DC energy minimization cycles, the DC energy remained stable, no sign of a variational collapse or prolapse was observed during the computations, that provides a numerical test (at least for the studied low-$Z$ range) of the current procedure.
For these reasons, we may say that the reported no-pair DC energies correspond to a basis set optimized to the non-relativistic energy. 
If further digits in the low-$Z$ or better results for the higher-$Z$ range are needed, then the DC optimizer will be further developed. 

We have also checked the singlet-triplet mixing by optimizing parameters with relaxing the non-relativistic spatial symmetry, as well as by an explicit $LS$ coupling of spatial functions optimized within their non-relativistic symmetry block and included with the appropriate spin function in the relativistic computation. The triplet contributions were negligibly small for the ground states of the studied low-$Z$ systems. Further details will be reported in future work.

\section{Positive-energy projection \label{ch:posenerproj}}

\subsection{Theoretical aspects and general concepts for the implementation \label{sec:ProjTheory}}
Due to the Brown--Ravenhall (BR) disease \cite{brownInteractionTwoElectrons1951}, the negative-energy states of the DC Hamiltonian are eliminated using a projection technique \cite{sucherFoundationsRelativisticTheory1980}. The fundamental idea of the procedure is based on the solution of the eigenvalue equation of a Hamiltonian without  particle-particle (electron-electron) interactions. Without these interactions, the negative- and positive-energy states of the different electrons are not coupled.
So, in the non-interacting case, the positive-energy states can be, in principle, selected and used to construct the ${\Lambda}^+$ projector as 
\begin{align}
  {\Lambda}^{+\left[ 16 \right]} 
  =  
  \sum_{\mu \in E^+} \left| \Psi_{0,\mu}^{\left(16 \right)} \right \rangle \left \langle  \tilde{\Psi}_{0,\mu}^{\left(16 \right)} \right| \ .\label{lambdadef}
\end{align}
$E^+$ is used to label the physically relevant, `positive-energy' space. In general, $\left \langle \tilde{\Psi}_{0,\mu}^{\left(16 \right)}\right | $ is the left and $\left | \Psi_{0,\mu}^{\left(16 \right)} \right \rangle$ is the right eigenvector of the $\mu$th state without particle-particle interactions, 
and the `0' subindex is used to emphasize that these are non-interacting states. 
It is necessary to distinguish the left and the right eigenvectors, if the underlying non-interacting Hamiltonian is not Hermitian (Sec.~\ref{ch:complexproj}).

We solve the eigenvalue equation of the Dirac--Coulomb Hamiltonian projected onto the positive-energy subspace, 
\begin{align}
    \bar{\mathcal{H}}_\mathrm{DC}^{\left[ 16 \right]}= \Lambda^{+\left[ 16 \right]} \mathcal{H}_\mathrm{DC}^{\left[ 16 \right]} \Lambda^{+\left[ 16 \right]} \ ,  
\end{align}
where $\bar{H}_\mathrm{DC}^{\left[ 16 \right]}$ is the so-called `no-pair' Hamiltonian \cite{sucherFoundationsRelativisticTheory1980,dyallIntroductionRelativisticQuantum2007,reiherRelativisticQuantumChemistry2015}.  

In order to determine the relativistic energies and wave functions, the matrix representation of the `no-pair' Hamiltonian is constructed over the positive-energy subspace,
\begin{align}
        \bar{\mathcal{H}}_{\mathrm{DC},\mu \nu}^{\left[ 16 \right]} =  \left \langle  \tilde{\Psi}_{0,\mu}^{\left(16 \right)} \right| \Lambda^{+\left[ 16 \right]} \mathcal{H}_\mathrm{DC}^{\left[ 16 \right]} \Lambda^{+\left[ 16 \right]}  \left| \Psi_{0,\nu}^{\left(16 \right)} \right \rangle=\left \langle  \tilde{\Psi}_{0,\mu}^{\left(16 \right)} \right| \mathcal{H}_\mathrm{DC}^{\left[ 16 \right]}  \left| \Psi_{0,\nu}^{\left(16 \right)} \right \rangle \ .\label{HDCmunu}
\end{align}
After the second equation in Eq.~(\ref{HDCmunu}), the $\Lambda^{+[16]}$ projectors have been suppressed, since the matrix representation is constructed over the positive-energy eigenfunctions of the non-interacting Hamiltonian. We also note that $\Lambda^{+[16]}$ was defined with these non-interacting (left- and right-) eigenfunctions, \rref{lambdadef}, for which the (bi)orthogonal property applies  
\begin{align}
\left \langle  \tilde{\Psi}_{0,\mu}^{\left(16 \right)} \right| \left . \Psi_{0,\nu}^{\left(16 \right)} \right \rangle = \delta_{\mu \nu} 1^{\left[16 \right]}  \; ,
\end{align}
where the left and right eigenvectors are understood to be normalized to each other in order to simplify the expressions in \rref{HDCmunu}.

During the numerical computations, first,  the non-interacting problem is solved in the ECG basis.
The (left and right) eigenvectors are written as
\begin{align}
  \left | \Psi_{0,\nu}^{\left(16 \right)} \right \rangle 
  &= 
  \sum_{i=1}^{N_\text{b}} \sum_{p=1}^{16} 
     c_{\nu, ip}^0 \mathcal{B}^{[16]} \, 1_p^{(16)} \left | \Theta_i \right \rangle  \ , \\
  \left \langle  \tilde{\Psi}_{0,\mu}^{\left(16 \right) } \right | 
  &= 
  \sum_{j=1}^{N_\text{b}} \sum_{p=1}^{16}
  \tilde{c}_{\mu, jp}^0  \left \langle \Theta_j \right |  {1_p^{(16)}}^T  \mathcal{B}^{[16]} \ ,
\end{align}
and they are substituted into \rref{HDCmunu} to obtain the matrix representation for the no-pair DC Hamiltonian:
\begin{align}
     \bar{H}_{\mathrm{DC},\mu \nu}^{\left[ 16 \right]} =& \sum_{i,j=1}^{N_\text{b}} \sum_{p,q=1}^{16} \tilde{c}_{\mu, jq}^{0} c_{\nu, ip}^{0} 
     %\nonumber  \\
     %& \cdot 
     \left \langle \Theta_j \left| 1_q^{(16)T} \mathcal{B}^{(16)} \mathcal{H}_\mathrm{DC}^{\left[ 16 \right]} \mathcal{B}^{(16)} 1_p^{(16)}\right| \Theta_i \right \rangle \ . \label{eq:projworkeq}
\end{align}
The matrix element $\left \langle \Theta_j \left| 1_q^{(16)T} \mathcal{B}^{(16)} \mathcal{H}_\mathrm{DC}^{\left[ 16 \right]} \mathcal{B}^{(16)} 1_p^{(16)}  \right| \Theta_i \right \rangle$ is calculated analytically and the permutation and spatial symmetries are considered according to Secs. \ref{sec:permutsym} and \ref{sec:spatsym}. The energy and the wave function are obtained by diagonalization of the $\bar{\bos{H}}_{\mathrm{DC}}^{\left[ 16 \right]} \in\mathbb{C}^{16N_\text{b}\times 16N_\text{b}}$ matrix. 

In short, the algorithm for computing the no-pair DC energies is as follows:
\begin{enumerate}
    \item build and diagonalize $\bos{H}_\mathrm{DC}^{\left[ 16 \right]}$ without the particle-particle interaction and using the ECG basis set,
    \item select the $E^+$-states,
    \item build $ \bar{\bos{H}}_{\mathrm{DC}}^{\left[ 16 \right]}$ with particle-particle interaction in the $E^+$-subspace using \rref{eq:projworkeq},
    \item diagonalize $ \bar{\bos{H}}_{\mathrm{DC}}^{\left[ 16 \right]}$. 
\end{enumerate}

\subsection{Algorithms for the construction of the positive-energy projector in an explicitly correlated basis}
\subsubsection{Projection with the complex-scaling technique \label{ch:complexproj}}
In this section, we adopt the complex-scaling approach, 
as it was originally proposed by Bylicki, Pestka, and Karwowski \cite{bylickiRelativisticHylleraasConfigurationinteraction2008} for defining a positive-energy projector, 
and generalize it to molecules.
First of all, similarly to resonance computations \cite{balslevSpectralPropertiesManybody1971,simonResonancesNBodyQuantum1973,moiseyevNonHermitianQuantumMechanics2011,jagauExtendingQuantumChemistry2017}, complex scaling is employed for the particle coordinates, 
\begin{align}
    r_i \rightarrow r_i \eem^{\iim \theta} \ ,
    \label{eq:ccrot}
\end{align}
where $\theta$ is a real parameter.
For atoms \cite{bylickiRelativisticHylleraasConfigurationinteraction2008}, the 
terms of the Dirac--Coulomb Hamiltonian are rescaled according to the following relations
\begin{align}
    \sigma_i \pdv{}{x_i} &\rightarrow \sigma_i \pdv{}{x_i}  \eem^{-\iim \theta} \ , \\
    \frac{1}{\sqrt{\br^2}} & \rightarrow  \frac{\eem^{-\iim\theta}}{\sqrt{\br^2}} \ , \\
    \frac{1}{\sqrt{\left(\br_i-\br_j\right)^2}} & \rightarrow   \frac{\eem^{-\iim\theta}}{\sqrt{\left(\br_i-\br_j\right)^2}} \ .
\end{align}

For molecules, some further considerations are necessary.
If the nuclear coordinate is not at the origin, the electron-nucleus interaction operator is not dilatation analytic, \emph{i.e.,}
\begin{align}
    \frac{1}{\sqrt{\left( \br_i - \bR_I \right)^2}} & \rightarrow \frac{\eem^{-\iim \theta} }{\sqrt{\left( \br_i - \bR_I \eem^{-\iim \theta}  \right)^2}} \ , \label{VneCCR}
\end{align}
and complex integration is required to evaluate the matrix elements. Since the analytic integral expressions are known for $\theta=0$, 
the expressions can be analytically continued for $\theta\neq0$, 
similarly to the non-relativistic computations reported in Refs.~ \cite{moiseyevAutoionizingStatesUsing1979,mccurdyComplexcoordinateCalculationMatrix1980,moiseyevNonHermitianQuantumMechanics2011}. Analytic continuation means in this case that the $\bos{R}_I$ nuclear coordinates are replaced with $\bos{R}_I \eem^{-\iim \theta}$ in the analytic integral expression. 
To carry out these types of non-dilatation analytic computations, we have generalized our original (real) ECG integral routines to complex arithmetic to be able to use complex-valued nuclear position vectors.

We have also tested an alternative (at a first sight na\"ive or `wrong') technique in which the nuclear coordinates are scaled together with the electronic coordinates,
\begin{align}
    \frac{1}{\sqrt{\left( \br_i - \bR_I \right)^2}} & \rightarrow \frac{\eem^{-\iim \theta} }{\sqrt{\left( \br_i - \bR_I  \right)^2}} \ , \label{VneCCRda}
\end{align}
that obviates the need of using complex arithmetic in the integral routines. But is this a correct procedure? Well, Moiseyev provides theoretical foundations for this approach  \cite{moiseyevNonHermitianQuantumMechanics2011} within a perturbative framework.
Since, we have `incorrectly' rotated also the fixed nuclear coordinates, we need to consider a (perturbative) series expansion for the `back rotation' of the nuclear coordinates to the real axis (where they should be, since they are treated in this work as fixed external charges)  \cite{moiseyevNonHermitianQuantumMechanics2011},
\begin{align}
  &E(\bos{R} \eem^{-i\theta}) \nonumber \\
  &= 
  E( \bos{R})  +(\eem^{\iim \theta}-1) \sum_I \bos{R}_I \frac{\partial E(\bos{R}_I)}{\partial \bos{R}_A}  + \dots \label{EnergCCRexp} 
  \\
  &\approx 
  E( \bos{R})  
  +
  \iim\theta
  \sum_I \bos{R}_I \frac{\partial E(\bos{R}_I)}{\partial \bos{R}_A} 
  -
  \theta^2
  \sum_I \bos{R}_I \frac{\partial E(\bos{I})}{\partial \bos{R}_A}  + 
  \dots \label{EnergCCRexp2}   
\end{align}
Thereby, we may use a dilatation analytic approach with \rref{VneCCRda}, but then, the resulting energy must be corrected, Eqs.~(\ref{EnergCCRexp})--(\ref{EnergCCRexp2}), by accounting for the necessary `back rotation' of the nuclei. 
At this point, it is important to emphasize that we are interested in the computation of bound states (so, not resonances!), and the complex coordinate scaling is used only to select the positive-energy states computed in an explicitly correlated (non-separable) basis. So, $\theta$ can take `any' (small) value that is already sufficiently large to distinguish the positive-energy states from the Brown--Ravenhall (electron-positron) states. In practical computations, $\theta$ is typically on the order of $\sim 10^{-5}$, and the `perturbative' correction due to the `back rotation' to the real energy is proportional to $\theta^2 \sim 10^{-10}$ that gives a (typically) negligible contribution. 

For these reasons (and in full agreement with the \emph{a posteriori} analysis of our computational results, Sec.~\ref{sec:results}), 
we have considered only the zeroth-order term in \rref{EnergCCRexp} and the $\theta \rightarrow 0$ limit. 

If we wanted to compute a resonance state, 
we cannot consider, in general, the small $\theta$ limit, since
a specific finite $\theta$ value corresponds to the stabilization of the (complex) resonance energy \cite{moiseyevNonHermitianQuantumMechanics2011}, and in that case, (higher-order) correction terms of the series expansion, \rrefsb{EnergCCRexp}, would be necessary to obtain good results. 
So, for resonance computations, the non-dilatation-analytic route, \rref{VneCCR}, appears to be the more practical choice, although the dilation-analytic approach is also acceptable, in principle.

Using \rrefsa{eq:dcmx}--\rrefsb{Wss}, the complex-scaled matrix element for two electrons can be written in the following form (we note that the kinetic balance, \rref{eq:kineticBalance}, is related to the basis set, and so, it is not included in the complex scaling),
 \begin{widetext}
\begin{align}
     &\langle \Phi_{jp}'^{(16)} | \mathcal{H}_{\mathrm{DC},\theta}^{[16]} | \Phi_i^{(16)} \rangle = {1^{(16)}_p}^T \langle \Theta_j | {B^{[16]}}^\dagger \mathcal{H}_{\mathrm{DC},\theta}^{[16]} B^{[16]} | \Theta_i \rangle d_i^{(16)} = \label{eq:dcmxt}\\
     & {1^{(16)}_p}^T  \left(%
       \begin{array}{cccc}
          \left \langle \Theta_j \left| W_{\mathrm{ll},\theta}^{[4]}  \right| \Theta_i \right \rangle & 
          \frac{\eem^{-\iim \theta}}{2m_2}\langle \Theta_j \left| p_2^2 \right | \Theta_i \rangle 1^{[4]}  & 
          \frac{\eem^{-\iim \theta}}{2m_1}\langle \Theta_j \left| p_1^2  \right | \Theta_i \rangle 1^{[4]} & 
          0^{[4]}  \\
          \frac{\eem^{-\iim \theta}}{2m_2}\langle \Theta_j \left| p_2^2 \right | \Theta_i \rangle 1^{[4]} &  
           \left \langle \Theta_j \left| W_{\mathrm{ls},\theta}^{[4]}  \right| \Theta_i \right \rangle & 
          0^{[4]} & 
          \frac{\eem^{-\iim \theta}}{8m_1m_2^2c^2}\langle \Theta_j \left| p_1^2 p_2^2 \right | \Theta_i \rangle 1^{[4]} \\
          \frac{\eem^{-\iim \theta}}{2m_1} \langle \Theta_i \left| p_1^2 \right | \Theta_j \rangle 1^{[4]}  &  
          0^{[4]} & 
          \left \langle \Theta_j \left| W_{\mathrm{sl},\theta}^{[4]}  \right| \Theta_i \right \rangle  & 
          \frac{\eem^{-\iim \theta}}{8m_1^2m_2c^2} \langle \Theta_i \left| p_1^2 p_2^2 \right | \Theta_j \rangle 1^{[4]}  \\
          0^{[4]}  & 
          \frac{\eem^{-\iim \theta}}{8m_1m_2^2c^2}\langle \Theta_i \left| p_1^2 p_2^2 \right | \Theta_j \rangle 1^{[4]}  &  
          \frac{\eem^{-\iim \theta}}{8m_1^2m_2c^2}\langle \Theta_i \left|p_1^2  p_2^2 \right | \Theta_j \rangle 1^{[4]}  & 
           \left \langle \Theta_j \left| W_{\mathrm{ss},\theta}^{[4]}  \right| \Theta_i \right \rangle
       \end{array}
     \right) d_i^{(16)} \ ,  \nonumber
     %\label{eq:phiiphij}
     \nonumber\\  
       &W_{\mathrm{ll},\theta}^{[4]}=  \left( \eem^{-\iim \theta}V +U_{\theta} \right) 1^{[4]}  \ ,  \\
      &W_{\mathrm{ls},\theta}^{[4]}=\frac{1}{4m_2^2c^2} \sum_{i,j=1}^3   p_{2i} \left(\eem^{-\iim \theta} V+ U_{\theta}\right) p_{2j}  \sigma_{2i}^{[4]} \sigma_{2j}^{[4]}  -  \frac{1}{2m_2} p_2^2  1^{[4]}  \ ,  \\
      & W_{\mathrm{sl},\theta}^{[4]}=\frac{1}{4m_1^2c^2} \sum_{i,j=1}^3 p_{1i} \left(\eem^{-\iim \theta} V+ U_{\theta}\right) p_{1j} \sigma_{1i}^{[4]}  \sigma_{1j}^{[4]}  -  \frac{1}{2m_1} p_1^2  1^{[4]}  \ ,  \\
     & W_{\mathrm{ss},\theta}^{[4]}=\frac{1}{16m_1^2m_2^2c^4}\sum_{i,j,k,l=1}^3   p_{1i} p_{2j}  \left(\eem^{-\iim \theta} V+ U_{\theta}\right) p_{2k} p_{1l} \sigma_{1i}^{[4]} \sigma_{1l}^{[4]}  \sigma_{2j}^{[4]} \sigma_{2k}^{[4]}   -\frac{(m_1+m_2)}{8m_1^2 m_2^2 c^2}   p_1^2 p_2^2  1^{[4]}  \ .  \label{Wsst}
\end{align}
\end{widetext}
If $U_{\theta}$ 
is treated as a non-dilation analytic operator, it is transformed according to \rref{VneCCR}. If it is used in a dilation analytic procedure, then  $U_{\theta}= \eem^{-\iim \theta} U$, \rref{VneCCRda}. We also note that Eqs.~(\ref{eq:dcmxt})--(\ref{Wsst}) recover \rrefsa{eq:dcmx}--\rrefsb{Wss} for $\theta=0$.

\begin{figure}
  \begin{center}
    \includegraphics[scale=1.]{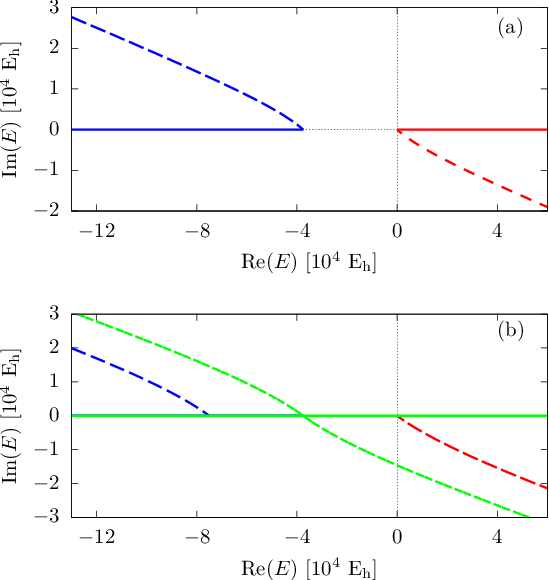}    
  \end{center}
    \caption{Visualization of the complex scaling technique to select the positive-energy states of the non-interacting two-electron system: (a) unbound single-particle energies for $\theta=0$ (solid) and for some finite $\theta>0$ (dashed) value; 
    (b) unbound non-interacting two-particle energies for the same angles. 
    The red, green, and blue colors indicate the electron-electron, electron-positron (or Brown--Ravenhall), and positron-positron states, respectively.
    \label{fig:H2CCR}}
\end{figure}

In what follows, we explain in detail how the complex coordinate rotation affects the non-interacting DC spectrum. Figure~\ref{fig:H2CCR} highlights an essential feature of the complex scaled DC energies that makes it possible to unambiguously identify and select the positive-energy states.
Figure~\ref{fig:H2CCR}a visualizes the energies of the unbound states for a single particle (either free or in interaction with a fixed, positive point charge with a charge number, $Z<\frac{\sqrt{3}}{2\alpha}$ \cite{lewinSpectralPollutionHow2010}). 
In the single-electron spectrum, 
the  `positronic' (negative-energy) and `electronic' (positive-energy) parts can be clearly identified, since they are separated by a finite energy gap.
If we apply the complex coordinate scaling, \rref{eq:ccrot}, the unbound positron and electron states rotate about different energy `centers' in the complex plane \cite{sebaComplexScalingMethod1988} and they are visualized by the dashed (blue and red) curves in the figure. 

If we use a non-separable (explicitly correlated) basis, we also need to consider the behaviour of two non-interacting particles (Fig.~\ref{fig:H2CCR}b). In such a basis, we can compute only the two-electron states (of the non-interacting system). The non-interacting, two-electron energy is the sum of the one-electron energies, but the one-electron energies are not obtained explicitly in the computation.
Due to the presence of the continuum both in the positronic and in the electronic parts, the two-electron energy spectrum covers the entire real axis (green, solid line in Fig.~\ref{fig:H2CCR}b).
Then, we consider the effect of the complex scaling. For `any' finite $\theta$ angle, three  branches appear in the complex plane depending on the sign of the one-particle contributions to the (non-interacting) two-particle energy. According to the different branches in the complex energy plane, the 
electron-electron (positive-energy), 
electron-positron (also called Brown--Ravenhall, BR),
and 
positron-positron states can be identified \cite{pestkaComplexCoordinateRotation2007,bylickiRelativisticHylleraasConfigurationinteraction2008}. 

In the next step, the Hamiltonian matrix for the system with interactions is constructed for the $\theta$ value for which the positive-energy states had been identified and the positive-energy projection can be carried out. In the last step, the `no-pair' energies are obtained by diagonalization of the projected matrix.

Identification of the positive-energy states was automated  using the linear relationship  \cite{bylickiRelativisticHylleraasConfigurationinteraction2008} that can be derived based on the behaviour of the one- and two-electron energies in the complex plane upon the complex scaling of the coordinate, 
\begin{align}
  f[\mathrm{Re}(E)] 
    &= 
    %-\tan \left( \theta \right) \mathrm{Re}(E) + mc^2 \ ,
    -\tan\theta\ \mathrm{Re}(E) + mc^2 \ .
\end{align} 
The intercept is set to be halfway between
the centers of the positron-electron and electron-electron branches, while the slope is parallel with the asymptotic, large Re($E$), behaviour of the complex eigenvalues \cite{sebaComplexScalingMethod1988}. During the selection procedure, state $\mu$ is identified as a positive-energy state, if the $\mathrm{Im}(E_\mu) > f[\mathrm{Re}(E_\mu)]$ condition is satisfied.

The projected energy was examined for different $\theta$ values. In general, it was found to be insensitive to the precise value of the $\theta$ rotation angle, until $\theta$ remained small. For large $\theta$ values, finite basis set effects become important. For too small $\theta$, distinction of the three branches becomes problematic in the finite (double or quadruple) precision arithmetic. 
At the same time, the results were sensitive to the number of states included in the projector. So, in practice, the selection was carried out according to the linear $f[\mathrm{Re}(E)]$ relation with the
additional constraint to keep the number of positive energy states fixed for every angle.  

If we assume that the number of the one-particle positronic and electronic states is equal, the number of the electron-electron states can be fixed to the quarter of the total number of states, $4N_\text{b}$. 
We note that in practical computations, the selection of a few points was not entirely unambiguous based on the $f[\mathrm{Re}(E)]$ linear relation and this additional `constraint' implemented in the automated procedure was necessary to have consistent results.

In the algorithm, we first sorted the states according to the distance of their (complex) energy from the $f[\mathrm{Re}(E)]$ line, and then, we kept the first $4N_\text{b}$ states in this list to define the positive-energy space.

\subsubsection{Projection with `cutting' in the energy}
For small nuclear charges, the relativistic `effects' are `small' in comparison with the non-relativistic energy. In this case, the positive-, the BR, and the negative-energy regions of the spectrum are well separated in a non-interacting, finite basis computation.
In this case, the positive-energy space can be selected in practice from the non-interacting,  two-electron computation by retaining the non-interacting states with (real) energies (without any complex scaling) that are larger than an estimated energy threshold. We will refer to this projection technique as `cutting' (in the energy spectrum).

The deficiency of this simple approach becomes apparent, if the nuclear charge and/or the number of basis functions is increased. Then, there  are electron-positron states that contaminate the electron-electron part of the space, and the corresponding BR states may appear as the lowest state of the (imperfectly) projected Hamiltonian.
At the same time, we note that this  simple `cutting' projector worked remarkably well for most of the low-$Z$ systems studied in this and in the forthcoming paper `Paper~II' \cite{ferencBreitInteractionExplicitly2021}.
The cutting projector has the advantage that it uses only real integral routines, and due to the hermiticity of the problem, the energy (and the underlying basis set) can be optimized variationally (until contaminant states appear). We used the `cutting' projector for exploratory computations and we have checked the final results using the, in principle, rigorous CCR projection procedure (that confirmed all significant digits of the cutting projection for the studied low-$Z$ systems in this work and in Refs.~\cite{jeszenszkiAllorderExplicitlyCorrelated2021,ferencBreitInteractionExplicitly2021}).

\subsubsection{Projection with `punching' in the energy list}
Since excellent results can be obtained with very small CCR angles (Tables~\ref{tab:heCCR} and \ref{tab:h2CCR}), we have experimented with discarding the states in the $\theta\rightarrow 0$ limit (followed numerically) for every state labelled as a BR state in the CCR projection procedure. In practice, we have discarded all states that have an energy less than an energy threshold and we `punched out' all higher-energy states from the positive-energy list that corresponded to a state identified as a BR state in the CCR procedure. This `punching' projector ideally combines the good features of cutting (hermiticity) and CCR (rigorous identification of the BR states). Its numerical performance and any disadvantages can be explored for higher-$Z$ systems (where the simple `cutting' projector fails) and is left for future work.

\subsubsection{An attempt to perform projection based on a determinant expansion}
In the previous subsection, the positive-energy states were selected directly from the solution of the non-interacting Hamiltonian using ECG basis functions. 
At the same time, a non-separable, explicitly correlated basis set is not an ideal choice for describing the non-interacting system that is not correlated. States of the non-interacting system are more naturally represented by anti-symmetrized products of one-particle functions, \emph{i.e., determinants}, in which, apart from the permutational anti-symmetrization there is no correlation between the electrons.
Of course, using determinants for describing interacting electrons leads to a less accurate representation in comparison with an explicitly correlated basis. For this reason, it appears to be necessary to use two different basis sets, one for describing the interacting and another set for describing the non-interacting states. We outline some early attempts to combine the two worlds, but without reporting any numerical results, because we think that some further considerations will be necessary for an efficient approach.
The accuracy of this route is currently limited by the size of the determinant space \cite{liuPerspectivesRelativisticQuantum2012,liuRelativisticExplicitCorrelation2017}.  

Along this route, first, the single-electron Dirac Hamiltonian (corresponding to a fixed nuclear geometry) is solved using a floating Gaussian basis. Then, only those functions are retained, for which the one-particle energy belongs to the purely electronic (positive-energy) states, 
\begin{align}
    \left| \varphi_{i}^{+(4)} \right \rangle &= 
    \sum_{j=1}^{n_\text{b}} \sum_p^4 d_{jip+} \mathcal{B}^{[4]} \mathcal{S}_\mathcal{B}^{[4]}  1_p^{(4)} \left| \chi_{j} \right \rangle \ , \label{onepartexpand}
\end{align}
where $\left| \chi_{j} \right \rangle$ is the floating Gaussian function, and $n_\text{b}$ is the number of the one-particle basis functions, $\mathcal{B}^{[4]}$ is a metric tensor corresponding to the kinetic balance condition, \rref{onepartB}, and $\mathcal{S}_\mathcal{B}^{[4]}$ ensures the one-particle spatial symmetry, \rrefsa{OBdef}--\rrefsb{eq:modrefsym}.
The two-particle determinant is constructed using the Tracy--Singh product as
\begin{align}
    \left| \Phi_{ij}^{\mathrm{D}+(16)} \right \rangle 
    =  
    \mathcal{A}^{[16]}  \left( \left| \varphi_{i}^{+(4)} \right \rangle \boxtimes \left| \varphi_{j}^{+(4)} \right \rangle \ \right) . \label{determincontruct}
\end{align}
Substituting \rref{onepartexpand} into \rref{determincontruct}, we can obtain the determinant (D) expansion in the Gaussian spinor-basis,
\begin{align}
    \left| \Phi_{ij}^{\mathrm{D}+(16)} \right \rangle  &=  \sum_{k,l=1}^{n_\text{b}} \sum_{p=1}^{16} d_{kl,ijp+} \mathcal{A}^{[16]} \mathcal{B}^{[16]} \mathcal{S}_\mathcal{B}^{[16]} 1_p^{(16)} \left|  X_{kl} \right \rangle \label{detansatz}\ ,
\end{align}
where the following notation was introduced:
\begin{align}
    \left|  X_{kl} \right \rangle &= \left| \chi_{k} \right \rangle \left| \chi_{l} \right \rangle \ , \\
            d_{kl,ijr+} &=   \sum_{p,q=1}^4 d_{kip+}  d_{ljq+} {1_r^{(16)}}^T  \left[ 1_p^{(4)} \boxtimes 1_q^{(4)} \right]\ .
\end{align}
Multiplying \rref{detansatz} with an ECG, \rref{spinoransatz}, from the left, we obtain
\begin{align}
    &\left\langle \Phi_{qp}^{(16)}  {\Big |} \Phi_{ij}^{\mathrm{D}+(16)} \right \rangle  = \label{expanddet} 
    %\sum_{kl} 
    \sum_{k,l=1}^{n_\text{b}} \sum_{q=1}^{16} 
    d_{kl,ijq+} {1_p^{(16)}}^T\underbrace{  \, \left\langle \Theta_q \left |  \mathcal{B}^{[16]} \mathcal{B}^{[16]} \mathcal{A}^{[16]} \mathcal{S}_\mathcal{B}^{[16]} \right|  X_{kl} \right \rangle}_{\tilde{S}_{q,kl}^{[16]}} 1_q^{(16)} \ . 
\end{align}
Then, we expand $\left|  \Phi_{ij}^{\mathrm{D}+(16)} \right \rangle$ in terms of the ECGs and multiply the expansion from the left with an ECG function, \rref{spinoransatz}, and obtain an alternative expression for \rref{expanddet}, 
\begin{align}
    &\left\langle \Phi_{qp}^{(16)}  {\Big |} \Phi_{ij}^{\mathrm{D}+(16)} \right \rangle  = \label{expandECG} 
    %\sum_{k} 
    \sum_{k=1}^{n_\text{b}} \sum_{q=1}^{16} c_{ij,kq}
    {1_p^{(16)}}^T \underbrace{  \, \left\langle \Theta_q \left |  \mathcal{B}^{[16]} \mathcal{B}^{[16]} \mathcal{A}^{[16]} \mathcal{S}_\mathcal{B}^{[16]} \right|  \Theta_k \right \rangle}_{S_{q,k}^{[16]}} 1_q^{(16)} \ . 
\end{align}
Since the right-hand sides of \rref{expanddet} and \rref{expandECG} are equal, we obtain the expansion coefficients as
\begin{align}
   c_{ij,mr} =  
   %\sum_p \sum_{kl} 
   \sum_{p=1}^{N_\text{b}} \sum_{k,l=1}^{n_\text{b}}  \sum_{q=1}^{16} 
  d_{kl,ijq+} {1_r^{(16)}}^T  \left( S^{[16]}\right)_{mp}^{-1} \bar{S}_{p,kl}^{[16]} 1_q^{(16)} \ . \label{cexpr}
\end{align}
Using these coefficients, we can define the positive-energy space for the ECG basis, while maintaining a Hermitian formulation and at the same time, in principle, accounting for explicit correlation necessary for the ppb convergence for the energy of the interacting system.  
Equation \rrefsb{expandECG} is critical in this approach that is exact only in the complete basis limit. 
Moreover, the spaces expanded by the determinants and the ECGs are not the same.
Calculating the rest of the space by including the negative-energy solutions in the ECG basis, we have found that it has a non-zero overlap with the positive-energy space. 
At the same time, this overlap disappears in the complete basis limit. Apparently, there is no unique way to treat this overlapping space. For exploratory computations (not reported in this work), we have used \rref{cexpr}, where the full overlapping space is part of the positive-energy space.

\clearpage
\section{Numerical results\label{sec:results}}
\subsection{Dirac--Coulomb energy for the helium atom \label{sec:heatom}}
\paragraph{Computational details}
The relativistic ground state of helium is dominated by $^1S^\text{e}$-type functions (where `e' is for even parity) that were generated by 
minimizing the non-relativistic ground-state energy. 
We used ECG functions, Eq.~(\ref{eq:ECGansatz}), with $\bos{s}=0$ to represent the spherical symmetry of the $S^\text{e}$-type functions.
For the largest basis set sizes, $N_\text{b}=300$ and 400,  the non-relativistic ground-state energy was converged within 1~n$\Eh$ of the reference value \cite{drakeHighPrecisionCalculations2006}.
The computations with these functions were made efficient by making use of the symmetry of the spatial ($\bos{s}=0$) and the singlet spin functions (further details are provided in the Supplementary Material). 
We have performed repeated refinement cycles of the basis function parameterization by minimization of the projected DC energy. Although, these refinement cycles improved the projected DC energy for small basis sets, for larger basis set sizes ($N_\text{b}>200$), the change was a fraction of a nE$_\mathrm{h}$.

Regarding the contribution of $^3P^\text{e}$-type functions, we have generated additional $\bos{s}\neq 0$ basis functions selected based on the minimization condition of the no-pair DC energy. The contributions of these functions was less than the current 1~n$\Eh$ convergence goal. Further details regarding the triplet contributions will be reported in future work.

In all computations, we used the speed of light as $c=\alpha^{-1}a_0\Eh/\hbar$ with 
$\alpha^{-1}=137.$035 999 084 \cite{Codata2018Recommended}.

\begin{table}[h]
  \caption{%
    Testing the positive-energy projection techniques for the Dirac--Coulomb energy of the ground state of the helium atom:
    comparison of the dilatation analytic ($E^\mathrm{da}_\DC$) and the energy cutting ($E^\mathrm{cutting}_\DC$) techniques.
    $N_\text{b}=300$ ECG basis functions are used with double-precision arithmetic.  
    All energy values are in $\Eh$. 
    \label{tab:heCCR}}
 \begin{tabular}{@{}l lc r @{}}
    \hline\hline\\[-0.35cm]
    & \multicolumn{3}{c}{proj = da($\theta$)} \\
    \cline{2-4} \\[-0.3cm]
    \multicolumn{1}{l}{$\theta$} &	
    \multicolumn{1}{c}{$\mathrm{Re} (E^\mathrm{proj}_\DC)$}  & \hspace{0.2cm} &
    \multicolumn{1}{c}{$\mathrm{Im} (E^\mathrm{proj}_\DC)$} \\
    \cline{1-4}\\[-0.3cm]
     0.000 000 01 & $-$2.903 856 630 628 & &  1.47 $\cdot 10^{-17}$ \\
     0.000 000 1  & $-$2.903 856 630 628 & &  1.47 $\cdot 10^{-16}$ \\
     0.000 001   & $-$2.903 856 630 628 & &  1.47 $\cdot 10^{-15}$ \\
     0.000 01    & $-$2.903 856 630 628 & &  1.47 $\cdot 10^{-14}$ \\
     0.000 1     & $-$2.903 856 630 628 & &  1.47 $\cdot 10^{-13}$\\
     0.001      & $-$2.903 856 630 628 & &  1.46 $\cdot 10^{-12}$ \\
     0.01       & $-$2.903 856 630 656 & &  1.61 $\cdot 10^{-11}$ \\
     0.1        & $-$2.903 856 632 538 & &  1.29 $\cdot 10^{-9}$ \hspace{0.02cm} \\
     0.2        & $-$2.903 856 632 442 & &  2.62 $\cdot 10^{-9}$ \hspace{0.02cm} \\
     0.5        & $-$2.903 856 823 509 & & $-$3.57 $\cdot 10^{-7}$ \hspace{0.02cm} \\
    \cline{1-4}\\[-0.3cm]
    proj = cutting:   &  $-$2.903 856 630 628	&	&  \multicolumn{1}{c}{0} \\
    \hline\hline
  \end{tabular}
\end{table}

\vspace{0.15cm}
\paragraph{Comparison of the energy cutting and the CCR projector}
During our work, we have noticed (in agreement with Karwowski and co-workers \cite{pestkaDiracCoulombHamiltonianNElectron2003,pestkaApplicationComplexcoordinateRotation2006,pestkaComplexCoordinateRotation2007}) that for small $Z$ values the positive-, the BR-, and the negative-energy non-interacting states are well separated. 
Hence, the simple `cutting' projector can be expected to work well. 

Table~\ref{tab:heCCR} shows the comparison of the projected DC energies obtained with the cutting and the dilatation analytic (da) CCR projector for several rotation angles. Regarding the CCR projector, it is important to note that 
the no-pair DC Hamiltonian is bounded from below (see also the more detailed and precise discussion in Sec.~III), and thus, we can aim to compute bound states. Hence, the complex coordinate rotation is used only to be able to distinguish the different non-interacting `branches'. Any rotation angle is appropriate (and gives the `same' numerical value for the bound-state energy) that is large enough---so, we can separate the non-interacting branches---, and at the same time, that is not too large---so, the finite basis set error remains small. For the present example (Table~\ref{tab:heCCR}), `small' means  $\theta\leq 0.01$. 
For the different `small' $\theta$ values, the imaginary part of the energy oscillates around 0 (and by using quadruple precision in the computations, we get numerical values closer to 0).
In all computations, the cutting projector could be unambiguously defined. The result of the energy-cutting procedure agrees to 10-11 digits with the da-CCR energy ($\theta\leq 0.01$). 

Of course, all results should be understood with respect to the no-pair Hamiltonian that is defined by the selected non-interacting states. Throughout this work, the non-interacting reference system is the one-electron problem in the field of the fixed nucleus (nuclei). We have tested the use of other reference systems. 
Their discussion is left for future work to be considered in relation with the QED (pair-)corrections, \emph{e.g.,} which reference system minimizes the corrections or which one offers the most practical option for the implementation of the corrections.

\vspace{0.5cm}
\paragraph{Comparison with literature data}
The no-pair DC energy obtained in this work for the helium ground state is $-2.903\ 856\ 631$~$\Eh$\ and it is considered to be converged on the order of 1~n$\Eh$ (see also Supplementary Material). The no-pair DC energy (corresponding to the same non-interacting model) reported by Bylicki, Pestka, and Karwowski \cite{bylickiRelativisticHylleraasConfigurationinteraction2008} is $-2.903\ 856\ 87$~$\Eh$. 
If all digits are significant for helium in Ref.~\cite{bylickiRelativisticHylleraasConfigurationinteraction2008}, then the following considerations can be made regarding the deviation from our no-pair DC energy.

On the one hand, Karwowski and co-workers considered the exact relativistic coalescence condition \cite{kutzelniggGeneralizationKatoCusp1989,liRelativisticExplicitCorrelation2012}, which is not accounted for in the ECG basis set used in this work. They report the effect of the singularity of the exact DC wave function at the coalescence point to be relevant for the 10th significant digit for the $Z>10$ range of heliumlike ions (the `effect' of the coalescence condition was reported only for this $Z$ range in the paper) \cite{pestkaGeminalsDiracCoulomb2012}. 

On the other hand, they  used an iterative kinetic balance condition that is different from the restricted balance used in the present work. Furthermore, they represented the large-large, large-small, and small-small subspaces by separate basis sets, and checked the quality of the representation for the non-relativistic limit by solving the Lévy--Leblond equation \cite{pestkaHylleraasCIApproachDiraccoulomb2003}.

In our work, the construction of the large-small model spaces and the kinetic balance condition ensured an exact fulfillment of the non-relativistic limit, although our kinetic balance condition can be considered as an approximation to the two-particle `relativistic'  kinetic balance condition of Ref.~\cite{simmenRelativisticKineticbalanceCondition2015} that is expected to provide a better representation of the negative-energy states than our balance. The effect of this difference for a positive-energy projected, \emph{i.e.,} no-pair computation remains to be explored in future work.

Furthermore, in a strict mathematical sense, neither the restricted, nor the relativistic \cite{simmenRelativisticKineticbalanceCondition2015}, nor the iterative \cite{pestkaApplicationComplexcoordinateRotation2006} kinetic balance conditions are complete, and only the use of the atomic balance would ensure a rigorous variational property in a strict mathematical sense \cite{lewinSpectralPollutionHow2010}. This also means for the projected (no-pair) energies of this work (also \cite{jeszenszkiAllorderExplicitlyCorrelated2021}) and of Ref.~\cite{bylickiRelativisticHylleraasConfigurationinteraction2008} that the positive-energy projectors constructed from the `quasi-variational' non-interacting computations may, in principle, contain a very small amount of negative-energy contamination (due to the approximate kinetic balances), resulting in slightly different `quasi-variational' energy values. It is necessary to note, however, that we have never experienced any sign of a variational collapse in our no-pair computations (that would be a clear indication of a negative-energy pollution). At the same time, `prolapse' may occur that would be difficult to notice, nevertheless, it might be responsible for the small deviation of the projected energy of this work and of Ref.~\cite{bylickiRelativisticHylleraasConfigurationinteraction2008}.
To better explore these aspects, we plan to test both the dual balance \cite{shabaevDualKineticBalance2004}, as well as the relativistic kinetic balance \cite{simmenRelativisticKineticbalanceCondition2015} techniques in future work. The implementation of the iterative kinetic balance approach of Karwowski and co-workers does not appear immediately straightforward for us. The application of the rigorous atomic balance does not seem to be feasible neither, in spite of a demonstration of its numerical applicability for a simple system \cite{dolbeaultEigenvaluesOperatorsGaps2000}.

Comparison of our no-pair DC energies with respect to precise energy values from the nrQED theory is provided in Sec.~\ref{sec:fwdc}.

\clearpage
\subsection{Dirac--Coulomb energy for the \texorpdfstring{$\mathrm{H}_2$}{} molecule \label{sec:h2mol}}

\begin{table}
  \caption{%
    Testing the positive-energy projection techniques for the Dirac--Coulomb energy of the H$_2$ molecule with fixed protons ($R_\text{pp}=1.4$~bohr):
    comparison of the dilatation analytic ($E^\mathrm{da}_\DC$), non-dilatation analytic ($E^\mathrm{nda}_\DC$), and energy cutting ($E^\mathrm{cutting}_\DC$) techniques.
    $N_\text{b}=700$ ECG basis functions are used with quadruple-precision arithmetic. All energy values are in $\Eh$.
  }
  \label{tab:h2CCR}
  \begin{tabular}{@{}llcc c  lccl @{}}
    \hline\hline\\[-0.35cm]
    &
    \multicolumn{3}{c}{proj = nda($\theta$)}  && 
    \multicolumn{3}{c}{proj = da($\theta$)}  & \\
    \cline{2-4}  \cline{6-8}\\[-0.3cm]
    \multicolumn{1}{c}{$\theta$} &	
    \multicolumn{1}{c}{$\mathrm{Re} (E^\mathrm{proj}_\DC)$}  & \hspace{0.cm} &
    \multicolumn{1}{c}{$\mathrm{Im} (E^\mathrm{proj}_\DC)$}   &&  
    \multicolumn{1}{c}{$\mathrm{Re} (E^\mathrm{proj}_\DC)$}  & \hspace{0.cm} &
    \multicolumn{1}{c}{$\mathrm{Im} (E^\mathrm{proj}_\DC)$}   & \\
    \cline{1-9}\\[-0.3cm]
    0.000 000 000 1	 & $-$1.174 489 753 666 & & 4.86 $\cdot 10^{-17}$  && $-$1.174 489 753 666 & &  7.14 $\cdot 10^{-11}$ & \\
    0.000 000 001	 & $-$1.174 489 753 666 & &  4.86 $\cdot 10^{-16}$  && $-$1.174 489 753 666 & & 7.14 $\cdot 10^{-10}$ & \\
    0.000 000 01	 & $-$1.174 489 753 666 & &  4.86 $\cdot 10^{-15}$ && $-$1.174 489 753 666 & & 	7.14 $\cdot 10^{-9}$ & \\
    0.000 000 1	 & $-$1.174 489 753 666 & &	4.86 $\cdot 10^{-14}$ && $-$1.174 489 753 666 & & 	7.14 $\cdot 10^{-8}$ & \\
    0.000 001	 &  $-$1.174 489 753 666 & & 4.86 $\cdot 10^{-13}$ && $-$1.174 489 753 666 & &  7.14 $\cdot 10^{-7}$ & \\
    0.000 01	     & 	$-$1.174 489 753 670 & &  4.86 $\cdot 10^{-12}$ && $-$1.174 489 753 667 & & 	7.14 $\cdot 10^{-6}$ & \\
     0.000 1	     & 	$-$1.174 489 754 085 & &	4.86 $\cdot 10^{-11}$ && $-$1.174 489 753 731 & & 7.14 $\cdot 10^{-5}$	 & \\
     0.001	     & 	$-$1.174 489 796 132 & &	3.93 $\cdot 10^{-11}$ && $-$1.174 489 760 167 & &  7.14 $\cdot 10^{-4}$ & \\
     0.002	     & 	$-$1.174 489 935 051 & &	 3.80 $\cdot 10^{-10}$  && $-$1.174 489 779 671 & & 1.43 $\cdot 10^{-3}$  & \\
     0.003	     & 	$-$1.174 489 319 219 & &	$-$2.21 $\cdot 10^{-8}$ \hspace{0.08cm}  && $-$1.174 489 812 176 & & 2.14 $\cdot 10^{-3}$ & \\
    \cline{1-9}\\[-0.3cm]
    proj = cutting: &  $-$1.174 489 753 666	&	&  \multicolumn{1}{c}{0} 	&& & & \\
    \hline\hline
  \end{tabular}
\end{table}

\paragraph{Computational details}
The no-pair DC ground state is dominated by the $^1\Sigma_\text{g}^+$ spin-spatial functions, and the contribution of triplet $^3\Pi_\text{g}$ and $^3\Sigma_\text{g}^-$-type functions is estimated to be smaller than the current 1~n$\Eh$ convergence goal. Further details will be reported in future work.
To carry out the computations with the dominant singlet $\Sigma_\text{g}^+$-type functions, we have fixed the $\bos{s}$ shift vectors of the ECGs on the interprotonic axis. 
In this case, the Hamiltonian matrix can be block diagonalized during the computation. 
The convergence of the energy with respect to the number of basis functions is shown in the Supplementary Material.

\vspace{0.15cm}
\paragraph{Comparison of the energy cutting, dilatation analytic CCR, and non-dilatation analytic CCR projectors}
The energies obtained from the non-dilatation analytic and dilatation analytic complex-scaling approaches are compared in Table~\ref{tab:h2CCR}. We note that it was necessary to use quadruple precision to observe the smooth behaviour at the sub-n$\Eh$ scale as it is shown in the table.
The non-dilatation analytic energies of H$_2$ (for various, small $\theta$ angles) have a (numerically) zero imaginary part, since they correspond to bound states. This behaviour is similar to the dilatation analytic computation of the helium atom. At the same time, we have carried out dilatation analytic computations also for the H$_2$ molecule that can be rigorously interpreted with a perturbative `back' rotation (Sec.~\ref{ch:complexproj}). It is interesting to note that the real parts of the energy of the non-dilatation analytic and dilatation analytic cases agree to 13 significant digits, but the imaginary parts are different. The imaginary part of the dilatation analytic energy is proportional to the $\theta$ rotation angle, and we would get the correct zero value if we employed the perturbative correction for the `back rotation', Eq.~(\ref{EnergCCRexp2}). We note that this perturbative correction for the real part of the energy is proportional to $\theta^2$ that is very small for small $\theta$ values. In this example, it was sufficient to use a $\theta$ value as small as $10^{-10}$ for the distinction of the positive-energy non-interacting branch from the BR and the negative-energy non-interacting states.

Both the dilatation analytic and the non-dilatation analytic energy agrees better than a ppb precision with the energy obtained by the energy `cutting' approach.
During the computations, we have observed some numerical sensitivity of the non-dilatation-analytic approach with respect to the $\theta$ rotation angle. This increased numerical uncertainty probably originates from the higher numerical uncertainty of the complex-valued Coulomb-type integrals. In our current implementation, the complex incomplete gamma function used in the complex Coulomb integrals can be evaluated only with 12 digits precision.

\clearpage
\vspace{0.15cm}
\subsection{Comparison of the variational no-pair Dirac--Coulomb energy with perturbative results\label{sec:fwdc}}

   \begin{table}
    \caption{%
    Comparison of the variational no-pair DC energies, $E_\DC^\proj$ in $\Eh$, computed in this work
    and the perturbative energies, $E_\DC^{(2)}$ and $E_\DC^{(3)}$ in $\Eh$, including the $\alpha^2$-order Foldy--Wouthuysen DC correction, $\varepsilon_\DC^\text{FW}$ \cite{dyallIntroductionRelativisticQuantum2007}, and the $\alpha^3$-order no-pair two-photon Coulomb correction, $\varepsilon_\text{CC}^{++}$ \cite{josephsucherEnergyLevelsTwoelectron1958}, for the example of the ground and the first excited states of the helium atom and for the ground state of the H$_2$ molecule ($R_\text{pp}=1.4$~bohr) and the H$_3^+$ molecular ion ($R_\text{pp}=1.65$~bohr).
    The deviation, $\delta = E^\proj_\DC - E_\DC^{(n)}$ ($n=2$ and 3), is shown in braces as $\lbrace \delta / \text{n}\Eh\rbrace$.
    The no-pair energies, $E_\DC^\proj$, are converged on the order of 2~n$\Eh$, further work with quadruple precision arithmetic will be necessary to confirm and improve upon this value. 
    \label{tab:fwdc}}
       \centering
       \begin{tabular}{@{}l@{\ \ \ \ \ }  rl rl @{}}
         \hline\hline
         & \multicolumn{2}{c}{He ($1\ ^1S_0$)} 
         & \multicolumn{2}{c}{He ($2\ ^1S_0$)} 
         \\
         \hline
         $E^\proj_\DC$
         & $-$2.903 856 631 &
         & $−$2.146 084 791 \\
         $E_\DC^{(2)} = E_\text{nr}+\alpha^2 \varepsilon^\text{FW}_\DC$ $^\text{a}$
         & $-$2.903 856 486 & $\lbrace -145\rbrace$
         & $-$2.146 084 769 & $\lbrace -22\rbrace$ \\
         $E_\DC^{(3)} = E_\text{nr}+\alpha^2 \varepsilon^\text{FW}_\DC + \alpha^3\varepsilon^{++}_\text{CC}$ $^\text{a}$
         & $-$2.903 856 620 & $\lbrace -11\rbrace$
         & $-$2.146 084 780 & $\lbrace -11\rbrace$ \\
         $\alpha^4\varepsilon_\text{non-rad}$ $^\text{b}$
         &  & $\lbrace -10.4\rbrace$ 
         &  & $\lbrace -11.2\rbrace$ \\[0.05cm]         
         \hline
         & \multicolumn{2}{c}{H$_2$} 
         & \multicolumn{2}{c}{H$_3^+$} \\
         \hline
         $E^\proj_\DC$
         & $-$1.174 489 754 & 
         & $-$1.343 850 527 & \\
         $E_\DC^{(2)} = E_\text{nr}+\alpha^2 \varepsilon^\text{FW}_\DC$ $^\text{a}$
         & $-$1.174 489 733 & $\lbrace -21\rbrace$
         & $-$1.343 850 503 & $\lbrace -24\rbrace$ \\
         $E_\DC^{(3)} = E_\text{nr}+\alpha^2 \varepsilon^\text{FW}_\DC + \alpha^3\varepsilon^{++}_\text{CC}$ $^\text{a}$
         & $-$1.174 489 754 & $\lbrace 0\rbrace$
         & $-$1.343 850 525 & $\lbrace -1\rbrace$ \\
         $\alpha^4\varepsilon_\text{non-rad}$ $^\text{b}$
         &  & $\lbrace -0.2\rbrace$ 
         &  & $\lbrace \text{n.a.}\rbrace$ \\         
        \hline\hline
       \end{tabular}
       \begin{flushleft}
         $^\text{a}$ %
         We used $\varepsilon^\text{FW}_\DC=-\frac{1}{8}\sum_{i=1}^N \langle\nabla_i^4\rangle
         + \frac{\pi}{2}\sum_{i=1}^N\sum_{A=1}^{\nnuc} Z_A \langle\delta(\br_{iA})\rangle
         - \pi \sum_{i=1}^N\sum_{j>i}^N \langle\delta(\br_{ij})\rangle$ \cite{dyallIntroductionRelativisticQuantum2007} 
        and
        $\varepsilon^{++}_\text{CC} = -\left(\frac{\pi}{2}+\frac{5}{3}\right)\langle\delta(\bos{r}_{ij})\rangle$ \cite{josephsucherEnergyLevelsTwoelectron1958}.
        The expectation values used to calculate the correction terms were taken from 
        Refs.~\cite{drakeHighPrecisionCalculations2006,drakeTheoreticalEnergiesStates1988,puchalskiCompleteAlpha6m2016,jeszenszkiInclusionCuspEffects2021}. \\
        $^\text{b}$ %
        The full non-radiative correction of order $\alpha^4$ for He (1 $^1S_0$ and 2 $^1S_0$)
        and H$_2$ (ground state, $R_\text{pp}=1.4$~bohr) were taken from Refs.~\cite{pachuckiAlphaMathcalRCorrections2006}
        and \cite{puchalskiCompleteAlpha6m2016}, respectively.
        (It is not evaluated for H$_3^+$ yet.)
       \end{flushleft}
   \end{table}

For the low-$Z$ range of atomic and molecular physics, the most precise computations including relativistic and QED corrections have been obtained by perturbative methods. For this reason, we compare the variational no-pair DC energy computed in this work with the perturbative values.
It was observed in Ref.~\cite{jeszenszkiAllorderExplicitlyCorrelated2021} that the variational and leading-order perturbative energies differ substantially already for the lowest $Z$ systems (see also $E_\DC^\proj$ \emph{vs.} $E_\DC^{(2)}$ in Table~\ref{tab:fwdc}). 

To better understand the origin of this difference, we consider the various physical contributions in the leading-order (non-radiative) QED corrections, $\oo{3}$, of the perturbative scheme first derived by Araki \cite{arakiQuantumElectrodynamicalCorrectionsEnergyLevels1957} and Sucher \cite{josephsucherEnergyLevelsTwoelectron1958}. Sucher~\cite{josephsucherEnergyLevelsTwoelectron1958} reported the contributing terms in detail, and the no-pair correction for the exchange of two Coulomb photons can be identified in his calculation that is present in the current variational treatment. 
Of course, the variational no-pair computation contains not only two-photon exchanges, but the full `Coulomb ladder' with the complete Dirac kinetic energy operator. Nevertheless, we may expect that the two-photon exchange is the most important, `leading-order' correction to the leading-order perturbative relativistic energy.

In Table~\ref{tab:fwdc}, we compare the no-pair variational and perturbative results for the singlet ground and first excited state of the helium atom, and 
for the ground electronic state of the H$_2$ and H$_3^+$ molecular systems near their equilibrium configuration. 
For H$_2$ and H$_3^+$, the inclusion of the $\oo{3}$ two-photon correction reduces the $-21$ and $-$24~n$\Eh$ deviation of the variational and perturbative energies to 0 and $-1$~n$\Eh$, respectively. 
Inclusion of the same correction in the perturbative energy of the 1~$^1S_0$ and 2~$^1S_0$ states of the helium atom reduces the  $-$145 and $-$22~n$\Eh$ deviations to $-$11~n$\Eh$ in both cases.

This remaining, still non-negligible difference of the variational and the perturbative DC energies indicate the importance of the perturbative corrections beyond the $\oo{3}$ level.
For the ground and the first excited singlet states of helium \cite{pachuckiAlphaMathcalRCorrections2006} and for the ground state of the H$_2$ molecule \cite{puchalskiCompleteAlpha6m2016}, the full $\oo{4}$ perturbative correction is also available. For this work, the non-radiative part of the correction is relevant that can be clearly identified in Refs.~\cite{pachuckiAlphaMathcalRCorrections2006,puchalskiCompleteAlpha6m2016}. Further analysis of the non-radiative perturbative corrections that would allow a direct comparison of the perturbative terms with our present no-pair energies is not available in Refs.~\cite{pachuckiAlphaMathcalRCorrections2006,puchalskiCompleteAlpha6m2016}. Nevertheless, we show the full non-radiative part of the $\oo{4}$ correction in Table~\ref{tab:fwdc} that allows a `rough' (probably appropriate in terms of an `order-of-magnitude') comparison with the remaining difference of the no-pair variational DC energy and the corresponding $\oo{3}$ perturbative value.
The good numerical agreement of the $E_\DC^\proj - E_\DC^{(3)}$ difference with $\alpha^4\varepsilon_\text{non-rad}$ in Table~\ref{tab:fwdc} is fortuitous and further developments are necessary for a direct comparison.
Furthermore, we think that the variational energies ($E_\DC^\proj$ reported in the table) are converged on the order of 1~n$\Eh$ for H$_2$, H$_3^+$ and for the $2\ ^1S_0$ state of helium, and on the order of 2~n$\Eh$ for the helium ground state ($1\ ^1S_0$).

\begin{figure}
    \centering
    \includegraphics[scale=0.7]{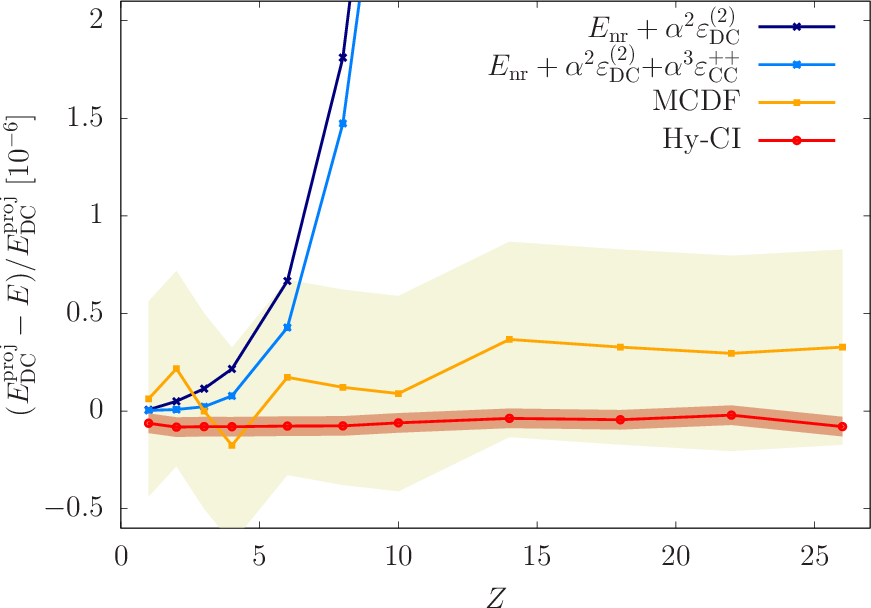}
    \caption{%
    Comparison of the variational no-pair DC energy with perturbative energies for two-electron ions (atom) with respect to the $Z$ nuclear charge number.
    The $E_\DC^\proj$ energy computed in this work (with a non-relativistically optimized %\madd{, 
    singlet
    %} \mdel{ECG} 
    basis set 
    %\mdel{, where the centers are fixed to the nucleus and only the singlet contribution are considered}}) 
    was used as reference and literature data is taken from Refs.~\cite{bylickiRelativisticHylleraasConfigurationinteraction2008} (Hy-CI) and \cite{PaGr90} (MCDF). 
    The perturbative corrections are calculated according to Ref.~\cite{josephsucherEnergyLevelsTwoelectron1958} using expectation values for the operators compiled from Ref.~\cite{drakeHighPrecisionCalculations2006} (up to $Z=4$) or computed in the present work ($Z>4$).
    (The data used to prepare this figure is provided in the Supplementary Material.)
    }
    \label{fig:fwdcZ}
\end{figure}

Figure~\ref{fig:fwdcZ} shows the comparison of the perturbative and no-pair variational DC energies for two-electron ions (atom), obtained in this work and in Refs.~\cite{bylickiRelativisticHylleraasConfigurationinteraction2008} and \cite{PaGr90},  over the $Z=1,...,26$ range of the nuclear charge number. 

Inclusion of the $\oo{3}$-order correction in the perturbative energy reduces the deviation for the lowest $Z$ values. Nevertheless, the variational-perturbative deviation rapidly increases and indicates the importance of the non-radiative `QED corrections' in the nrQED scheme beyond the $\oo{3}$ leading order already in the low $Z$ range.

\begin{figure}
    \centering
    \includegraphics[scale=0.7]{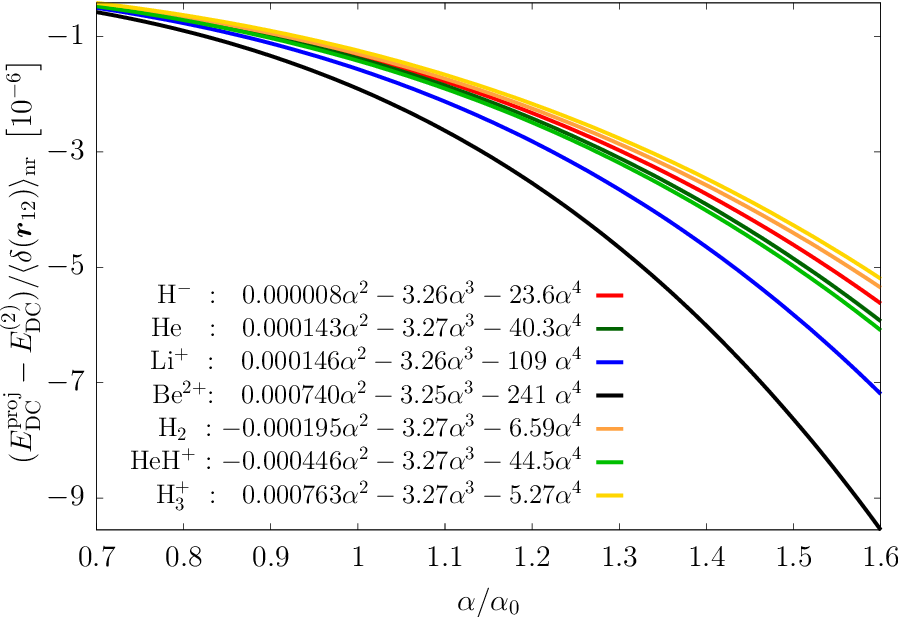}
    \caption{%
      Dependence of the no-pair DC energy, $E^\proj_\text{DC}$, on the value of the $\alpha$ coupling constant of the electromagnetic interaction. 
      We use Hartree atomic units and 
      $\alpha_0$ labels $1/137.$035~999~084 \cite{Codata2018Recommended}. 
      The data points, used for fitting the polynomials, were computed at the $\alpha=1/(\alpha_0+n)$, $n=-50, \dots, 50$ values.
      The $\alpha^2$, leading-order relativistic DC energy values, $E_\DC^{(2)}$, are compiled from 
      Refs.~\cite{drakeHighPrecisionCalculations2006,puchalskiRelativisticCorrectionsGround2017,jeszenszkiInclusionCuspEffects2021,jeszenszkiAllorderExplicitlyCorrelated2021} (and the contributions are collected in the Supplementary Material of Paper~II).
    \label{fig:alphascalingDC}
    }
\end{figure}

The no-pair DC energy can be computed for different $\alpha$ values that allows us to numerically determine its $\alpha$ dependence, and thereby, the `leading' and `higher-order' non-radiative QED corrections for the Coulomb interaction and positive-energy states. We have computed $E_\DC^\proj(\alpha)$ for 101 different $\alpha$ values distributed over the $\alpha\in[0.7,1.6]\alpha_0$ interval, where $\alpha_0=1/137.035 999 084$ is the CODATA18 recommended value \cite{Codata2018Recommended}. It was possible to fit (Fig.~\ref{fig:alphascalingDC}) the $\sum_{n=2}^m c_m \alpha^m$ polynomials to the data points with $m=3$ and 4. The $c_2$ and $c_3$ coefficients remained stable upon the inclusion of the $c_4\alpha^4$ term in the fitted function. It is interesting to note that $c_4>c_3$ for all systems studied. For the physically relevant $\alpha_0$ value, the $c_4\alpha^4$ fourth-order term is smaller than the $c_3\alpha^3$ third-order contribution. As $Z$ increases, the relative importance of the higher-order term increases: the fourth-order contribution is only 10~\%\ of the third-order term for helium, whereas it is already 50~\%\ for Be$^{2+}$.
The fitted functions shown in Fig.~\ref{fig:alphascalingDC} are `normalized' with $\langle\delta(\bos{r}_{12})\rangle_\text{nr}$ that brings the different systems  (helium-isoelectronic ions, H$_2$, H$_3^+$, and HeH$^+$) to the same scale in the figure. The third-order contribution is found to be 
\begin{align}
  c_3 \alpha^3 = -3.26(1)\ \langle\delta(\bos{r}_{12})\rangle_\text{nr}\ \alpha^3
\end{align}
in excellent agreement with Sucher's formal result, which can be found in Eq. (3.99) on p. 52 of Ref.~\cite{josephsucherEnergyLevelsTwoelectron1958}:
\begin{align}
  \varepsilon_\text{CC}^{++} \alpha^3
  =
  -
  \left(%
    \frac{\pi}{2}+\frac{5}{3}
  \right)\ 
  \langle\delta(\bos{r}_{12})\rangle_\text{nr}\ \alpha^3
  \approx 
  -3.24\ \langle\delta(\bos{r}_{12})\rangle_\text{nr}\ \alpha^3 \; .
\end{align}

\section{Summary and conclusion} \label{sec:conclusions}
Theoretical and algorithmic details are reported for an explicitly correlated, no-pair Dirac--Coulomb framework. Options for positive-energy projection techniques are considered in detail. The computed variational no-pair Dirac--Coulomb energies are compared with the corresponding energy values of the non-relativistic quantum electrodynamics framework for low $Z$ systems, and it is found that higher-order non-radiative QED corrections become increasingly important for an agreement better than $1:10^9$ beyond $Z=1$. Extension of the present framework with the Breit interaction is reported in the forthcoming paper \cite{ferencBreitInteractionExplicitly2021}.

\vspace{0.5cm}
\section*{Data availability statement}
The data that support findings of this study is included in the paper or in the Supplementary Material.

\vspace{0.5cm}
\section*{Supplementary Material}
The supplementary material contains (a) special matrix elements for spherically symmetric and singlet basis functions; (b) convergence of the computed energies; and (c) data for Figure~2.

\vspace{0.5cm}
\begin{acknowledgments}
\noindent Financial support of the European Research Council through a Starting Grant (No.~851421) is gratefully acknowledged. DF thanks a doctoral scholarship from
the ÚNKP-21-3 New National Excellence Program of the Ministry for Innovation and Technology from the source of the National Research, Development and Innovation Fund (ÚNKP-21-3-II-ELTE-41). 
We thank the Reviewers for their thoughtful comments that helped us to improve this work (Paper I \& II).
\end{acknowledgments}

\bibliography{book,ComplexScaling,Dirac-Coulomb,KineticBalance,LowerBound,MathAndConstants,NonAdiabatic,H2,programs,Relativistic,relatquantum,ExplicitGauss,ExplicitExponential,He,Hthreep,qed,QED,experiments,transcorrelated,theses}

\clearpage
%\input{DiracCoulomb_som_inc}

%%%%%%%%%%%%%%%%%%%%%%%%%%%%%%%%%%%%%%%%%%%%%%%%%%%%%%%%%%%%%%%%%%%%%%%%%
%%%%%%%%%%%%%%%%%%%%%%%%%%%%%%%%%%%%%%%%%%%%%%%%%%%%%%%%%%%%%%%%%%%%%%%%%
%%%%%%%%%%%%%%%%%%%%%%%%%%%%%%%%%%%%%%%%%%%%%%%%%%%%%%%%%%%%%%%%%%%%%%%%%
%%%%%%%%%%%%%%%%%%%%%%%%%%%%%%%%%%%%%%%%%%%%%%%%%%%%%%%%%%%%%%%%%%%%%%%%%
%%%%%%%%%%%%%%%%%%%%%%%%%%%%%%%%%%%%%%%%%%%%%%%%%%%%%%%%%%%%%%%%%%%%%%%%%
%%%%%%%%%%%%%%%%%%%%%%%%%%%%%%%%%%%%%%%%%%%%%%%%%%%%%%%%%%%%%%%%%%%%%%%%%
\newpage \renewcommand{\theequation}{1.\arabic{equation}} %
\setcounter{section}{0} \setcounter{equation}{0}

\setcounter{section}{0}
\renewcommand{\thesection}{S\arabic{section}}
\setcounter{subsection}{0}
\renewcommand{\thesubsection}{S\arabic{section}.\arabic{subsection}}

\setcounter{equation}{0}
\renewcommand{\theequation}{S\arabic{section}.\arabic{equation}}

\setcounter{table}{0}
\renewcommand{\thetable}{S\arabic{table}}

\setcounter{figure}{0}
\renewcommand{\thefigure}{S\arabic{figure}}

~\\[0.cm]
\begin{center}
\begin{minipage}{0.8\linewidth}
\centering
\textbf{Supplementary Material} \\[0.25cm]

\textbf{Variational Dirac--Coulomb explicitly correlated computations for atoms and molecules}
\end{minipage}
~\\[0.5cm]
\begin{minipage}{0.6\linewidth}
\centering

P\'eter Jeszenszki,$^1$  D\'avid Ferenc,$^1$ and Edit M\'atyus$^{1,\ast}$ \\[0.15cm]

$^1$~\emph{ELTE, Eötvös Loránd University, Institute of Chemistry, 
Pázmány Péter sétány 1/A, Budapest, H-1117, Hungary} \\[0.15cm]
$^\ast$ edit.matyus@ttk.elte.hu \\
\end{minipage}
~\\[0.15cm]
(Dated: January 4, 2022)
\end{center}

~\\[1cm]
\begin{center}
\begin{minipage}{0.8\linewidth}
\noindent %
Contents: \\
S1. Special matrix elements for spherically symmetric and singlet basis functions  \\
S2. Convergence of the computed energies  \\
S3. Data for Figure~2  \\
\end{minipage}
\end{center}

\clearpage
\section{Reduction of the Dirac--Coulomb matrix for spherically symmetric singlet functions \label{app:singlet}}
 For the special case of singlet spin and spherical spatial symmetry ($\bos{s}=0$ for the ECG basis, Eq.~(24)) of two-fermion systems, 
 we consider the sixteen-to-four-dimensional reduction of the Dirac--Coulomb matrix and the resulting four-dimensional block structure. Let us start with a sixteen-dimensional submatrix, Eqs.~(27)--(32).
 %\mdel{\rrefsa{eq:dcmx}--\rrefsb{Wss}. }
 %
 The overlap matrix is diagonal, and any spin mixing can arise only from the large-small, small-large, and small-small diagonal blocks of the Hamiltonian matrix. 
 
For spherically symmetric basis functions, all the blocks except for the small-small diagonal block in Eq.~(27) can be written as a scalar times a four-dimensional unit matrix. For the reduction of the the small-small block, further considerations are necessary.

The small-small matrix elements are
 \begin{align}
     &\left \langle \Theta_\mu \right| p_{1i} p_{2j} \left( V  +U \right)p_{2k} p_{1l} \left| \Theta_\nu \right \rangle = \label{intexpr} 
      \delta_{il}\delta_{jk} \alpha_{\mu \nu,ij}  \\
     &\hspace{1cm}+\delta_{ij} \delta_{kl}(1-\delta_{il})\beta_{\mu \nu,il}+ 
     \delta_{ik} \delta_{jl}(1-\delta_{il})\gamma_{\mu \nu,il} \nonumber \ 
\end{align}
with
\begin{align}
     &\alpha_{\mu \nu,ij} =  \int \mbox{d} \br \,  D^{\mu \mu}_{12,i}  D^{\nu \nu}_{12,j}  \Theta_\mu \left( \br \right) \left[ V\left(\br\right) +U\left(\br\right)\right]\Theta_\nu \left( \br \right) \ , \\
     &\beta_{\mu \nu,il} = \int \mbox{d} \br \,  D^{\mu \nu}_{12,i}  D^{\mu \nu}_{12,l}  \Theta_\mu \left( \br \right) \left[ V\left(\br\right) +U\left(\br\right)\right] \Theta_\nu \left( \br \right) \ , \\
     &\gamma_{\mu \nu,il}  = \int \mbox{d} \br \,  D^{\mu \nu}_{11,i}  D^{\mu \nu}_{22,l}  \Theta_\mu \left( \br \right) \left[ V\left(\br\right) +U\left(\br\right)\right] \Theta_\nu \left( \br \right) \ , \\
      &D^{\mu \nu}_{ab,i} =  \br^T \underline{\bA}_\mu \underline{\bE}_{a i,  b i} \underline{\bA}_\nu \br \ , \\
      &\underline{\bE}_{a i,  b j} = \delta_{ab} \delta_{ij} \ .
\end{align}
By considering the effect of the different $i,j,l$ spatial directions on the integral values, we find that
we need only four integrals to evaluate Eq.~(\ref{intexpr}) that are labelled as follows
\begin{align}
     \alpha'_{\mu \nu} &= \alpha_{\mu \nu, ii}   \ , \\
     \alpha_{\mu \nu} &= \alpha_{\mu \nu, ij} \ , \\
     \beta_{\mu \nu} &= \beta_{\mu \nu, ij} \ , \\
      \gamma_{\mu \nu} &= \gamma_{\mu \nu, ij} \ 
\end{align}
with $i,j=1,2,3$ and $i\ne j$. 
The multiplication of the sigma matrices have been simplified  by using the Dirac identity~\cite{diracQuantumTheoryElectron1928}, 
\begin{align}
  \left( \bos{a} \bos{\sigma}\right) \left(  \bos{b} \bos{\sigma} \right)= \left( \bf \bos{ab} \right)\bI + i \left( {\bf a} \times {\bf b} \right)  \bos{\sigma}  \label{sigmaexpr}
\end{align}
with $\bos{a},\bos{b} \in\mathbb{C}^3$. As a result, the product of the four $\sigma$ matrices in \rrefsa{intexpr} can be written as a product of two $\sigma$ matrices
 \begin{align}
     \sum_{i,j,k,l=1}^3 \sigma_{1i} \sigma_{2j} \sigma_{2k} \sigma_{1l} \left \langle \Theta_\mu \right| p_{1i} p_{2j} \left( V + U \right) p_{2k} p_{1l} \left| \Theta_\nu \right \rangle \ =  \label{foursigma}\\
    \left(3\alpha'_{\mu \nu}+6 \alpha_{\mu \nu} \right)\bI  + 2\left(\gamma_{\mu \nu}- \beta_{\mu \nu} \right)\sum_{i=1}^3 \sigma_{1i}\sigma_{2i} \ , \nonumber
 \end{align}
where
 \begin{align}
     \sum_{i=1}^3 \sigma_{1i}\sigma_{2i} = \left(
     \begin{array}{cccc}
     1 & 0 & 0 & 0 \\
     0 & -1 & 2 & 0 \\
     0 & 2 & -1 & 0 \\
     0 & 0 & 0 & 1
     \end{array} 
     \right) \ .
 \end{align}
 Thus, \rref{foursigma} can be written as
 \begin{widetext}
  \begin{align}
     &\sum_{i,j,k,l=1}^3 \sigma_{1i} \sigma_{2j} \sigma_{2k} \sigma_{1l} \left \langle \Phi_\mu \right| p_{1i} p_{2j} V p_{2k} p_{1l} \left| \Phi_\nu \right \rangle = \\
     & \nonumber \left(
     \begin{array}{cccc}
     3\alpha'_{\mu \nu}+6 \alpha_{\mu \nu} +2(\gamma_{\mu \nu}-\beta_{\mu \nu}) & 0 & 0 & 0 \\
     0 &  3\alpha'_{\mu \nu}+6\alpha_{\mu \nu} +2(\beta_{\mu \nu}-\gamma_{\mu \nu}) & 4(\gamma_{\mu \nu}-\beta_{\mu \nu}) & 0 \\
     0 & 4(\gamma_{\mu \nu}-\beta_{\mu \nu}) & 3\alpha'_{\mu \nu}+6\alpha_{\mu \nu} +2(\beta_{\mu \nu}-\gamma_{\mu \nu}) & 0 \\
     0 & 0 & 0 & 3\alpha'_{\mu \nu}+6\alpha_{\mu \nu} +2(\gamma_{\mu \nu}-\beta_{\mu \nu}) 
     \end{array} 
     \right) \ .
 \end{align}
 \end{widetext}
 The singlet matrix element can be obtained by choosing the anti-symmetric solution of two-dimensional block matrix along the diagonal, and it is $3\alpha'_{\mu \nu}+6(\alpha_{\mu \nu}+\beta_{\mu \nu}-\gamma_{\mu \nu})$.

\vspace{1cm}
The 16-to-4-dimensional reduction simplifies also the evaluation of the effect of the symmetry operators. 
For the spatial symmetry operators, the spinor part becomes a four-dimensional unit matrix. 

Regarding the anti-symmetrization of the wave function, the singlet spin component of the two-fermion systems is already antisymmetric to the particle exchange, hence, the `rest' of the operator must be symmetric 
for the permutation of the particles and can be written as
\begin{align}
     {\mathcal S}^{[4]}_+=1^{[4]}+\mathbb{P}^{\mathrm{ls}[4]}_{12} P_{12} \; ,
\end{align}
where $\mathbb{P}^{\mathrm{ls}[4]}_{12}$ describes the effect of the permutation on the large-small components, Eq.~(34).

\clearpage
\section{Convergence tests}
\begin{table}[h]
  \caption{%
    Convergence of the
    non-relativistic, $E_\text{nr}$, 
    non-projected, $E_{\DC}^\mathrm{bare}$, and 
    projected, $E_{\DC}^\mathrm{proj}$, Dirac--Coulomb energies of the ground state of the helium atom. For comparison, the convergence of the non-relativistic energy is also shown.
    All energies are in $\Eh$. 
    \label{tab:he}}
 \begin{tabular}{@{}l@{\ }l@{\ }l@{\ }l@{}}
    \hline\hline\\[-0.35cm]
    \multicolumn{1}{l}{$\nb$} &	
   \multicolumn{1}{c}{$E_\mathrm{nr}$}  &
    \multicolumn{1}{c}{$E_\DC^\mathrm{bare}$}   &
    \multicolumn{1}{c}{$E_{\DC}^\mathrm{proj}$}  \\
    \cline{1-4}\\[-0.3cm]
     50	 & 	$-$2.903 714 988 & 	$-$2.903 847 307    &	$-$2.903 847 334 \\
     100 &  $-$2.903 724 149 &	$-$2.903 856 311	&	$-$2.903 856 311 \\     
     150 &  $-$2.903 724 306 &	$-$2.903 856 487	&	$-$2.903 856 590 \\
     200 &  $-$2.903 724 368 &	$-$2.903 856 503	&	$-$2.903 856 622 \\
     300 &  $-$2.903 724 376 &	$-$2.903 856 504	&	$-$2.903 856 631 \\
     400 &  $-$2.903 724 377  &	$-$.903 856 505    &	$-$2.903 856 631 \\ 
    \cline{1-4}\\[-0.3cm]
    $E_\text{lit}$  & 
    $-$2.903 724 377 \cite{drakeHighPrecisionCalculations2006}	&
    $-$2.903 856 84 \cite{pestkaComplexCoordinateRotation2007} &	
    $-$2.903 856 87 \cite{bylickiRelativisticHylleraasConfigurationinteraction2008} \\
    $E_\text{lit}$  &    
      & $-$2.903 856 74 \cite{simmenRelativisticKineticbalanceCondition2015} 	&	\\
    \hline\hline
  \end{tabular}
\end{table}

\begin{table}[h]
  \caption{%
    Convergence of the
    non-relativistic, $E_\text{nr}$, 
    non-projected, $E_{\DC}^\mathrm{bare}$, and 
    projected, $E_{\DC}^\mathrm{proj}$, Dirac--Coulomb energies of 
    the ground state of the H$_2$ molecule with an $R_\text{pp}=1.4$~bohr interprotonic separation. 
    All energies are in $\Eh$.  
  \label{tab:h2}}
  \begin{tabular}{@{}l lc lc @{}}
    \hline\hline\\[-0.35cm]
    \multicolumn{1}{c}{$\nb$} &	
    \multicolumn{1}{c}{$E_\mathrm{nr}$}  &
    \multicolumn{1}{c}{$E_\DC^\text{bare}$}   &
    \multicolumn{1}{c}{$E_{\DC}^\text{proj}$} &  \\
    \cline{1-5}\\[-0.3cm]
     250	 & $-$1.174 475 698 & $-$1.174 489 721 & $-$1.174 489 738		& \\
     500	 & $-$1.174 475 713 & $-$1.174 489 734 & $-$1.174 489 753		& \\     
     700	 & $-$1.174 475 714 & $-$1.174 489 735 & $-$1.174 489 754		& \\
     1000    & $-$1.174 475 714 & $-$1.174 489 735 & $-$1.174 489 754	  & \\
     1200    & $-$1.174 475 714 & $-$1.174 489 735 & $-$1.174 489 754	  & \\
    \hline \\[-0.3cm]
     $E_\text{lit}$    & $-$1.174 475 714 \cite{puchalskiRelativisticCorrectionsGround2017} &  & 	  & \\    
    \hline\hline
  \end{tabular}
\end{table}

\clearpage
\section{Data for Figure~2}
\begin{table}[h]
    \centering
    \caption{%
      Comparison of the variational no-pair and the perturbative energies for the Dirac--Coulomb model of two-electron ions (atom) with an increasing $Z$ nuclear charge number.
      The perturbative corrections were calculated according to 
        $\varepsilon^\text{FW}_\DC=-\frac{1}{8}\sum_{i=1}^N \langle\nabla_i^4\rangle
         + \frac{\pi}{2}\sum_{i=1}^N\sum_{A=1}^{\nnuc} Z_A \langle\delta(\br_{iA})\rangle
         - \pi \sum_{i=1}^N\sum_{j>i}^N \langle\delta(\br_{ij})\rangle$ (e.g., \cite{josephsucherEnergyLevelsTwoelectron1958,dyallIntroductionRelativisticQuantum2007})
        and
        $\varepsilon^{++}_\text{CC} = -\left(\frac{\pi}{2}+\frac{5}{3}\right)\langle\delta(\bos{r}_{12})\rangle$ \cite{josephsucherEnergyLevelsTwoelectron1958}.
      }
    \label{tab:FWPTDCcompZ}
    \begin{tabular}{{@{}l  lr lr lr lr@{}}}
    \hline \hline
       $Z$  && \multicolumn{1}{c}{1} &\hspace{0.5cm}& \multicolumn{1}{c}{2} &\hspace{0.5cm}& \multicolumn{1}{c}{3} &\hspace{0.5cm}& \multicolumn{1}{c}{4} \\ \hline
       $E_\mathrm{DC}^\mathrm{proj}$ [this work]  
       && $-$0.527 756 733 && $-$2.903 856 631 && $-$7.280 698 899 && $-$13.658 257 603 \\
       $E_\mathrm{DC,Hy-CI}^\mathrm{proj}$ \cite{bylickiRelativisticHylleraasConfigurationinteraction2008}  && $-$0.527 756 766  && $-$2.903 856 87 \hspace{0.07cm} && $-$7.280 699 48 \hspace{0.07cm} && $-$13.658 258 7 $\hspace{0.22cm}$\\
       $E_\mathrm{MCDF}^\mathrm{proj}$ \cite{parpiaAccurateDiracCoulombEnergies1990}  && $-$0.527 756 7 \hspace{0.2cm} && $-$2.903 856 \hspace{0.5cm} && \multicolumn{1}{c}{(n.a.)} && $-$13.658 26  $\hspace{0.67cm}$ \\
       $E_\text{nr}$ $^\text{a}$ && $-$0.527 751 016 &$\hspace{0.5cm}$& $-$2.903 724 375 &$\hspace{0.5cm}$& $-$7.279 913 413 &$\hspace{0.5cm}$& $-$13.655 566 234 \\
       $E_\DC^{(2)} = E_\text{nr}+\alpha^2 \varepsilon^\text{FW}_\DC$ $^\text{a}$   && $-$0.527 756 730 && $-$2.903 856 486 && $-$7.280 698 064 && $-$13.658 254 651  \\
       $E_\DC^{(3)} = E_\text{nr}+\alpha^2 \varepsilon^\text{FW}_\DC + \alpha^3\varepsilon^{++}_\text{CC}$ $^\text{a}$  && $-$0.527 756 733 && $-$2.903 856 620 && $-$7.280 698 735  && $-$13.658 256 567  \\
       \hline \hline
    \end{tabular}
    ~\\[0.5cm]
    \begin{tabular}{{@{}l  lr lr lr lr@{}}}
    \hline \hline
       $Z$  && \multicolumn{1}{c}{6} &\hspace{0.5cm}& \multicolumn{1}{c}{8} &\hspace{0.5cm}& \multicolumn{1}{c}{10} &\hspace{0.5cm}& \multicolumn{1}{c}{14} \\ \hline
       $E_\mathrm{DC}^\mathrm{proj}$ [this work] && 
       $-$32.421 015 6 && $-$59.205 197 2 && $-$94.028 418 4 && $-$187.888 169 \\
       $E_\mathrm{DC,Hy-CI}^\mathrm{proj}$ \cite{bylickiRelativisticHylleraasConfigurationinteraction2008}  && $−$32.421 018 1  && $−$59.205 201 7  && $−$94.028 424 1  && $−$187.888 176 \\
       $E_\mathrm{MCDF}^\mathrm{proj}$ \cite{parpiaAccurateDiracCoulombEnergies1990}  && $−$32.421 01 \hspace{0.3cm} && $−$59.205 19  \hspace{0.3cm} && $−$94.028 41 \hspace{0.3cm} && $−$187.888 1  $\hspace{0.2cm}$ \\
       $E_\text{nr}$  $^\text{b}$ && $-$32.406 246 8 &$\hspace{0.5cm}$& $-$59.156 595 1 &$\hspace{0.5cm}$& $-$93.906 806 4 &$\hspace{0.5cm}$& $-$187.407 050 \\
       $E_\DC^{(2)} = E_\text{nr}+\alpha^2 \varepsilon^\text{FW}_\DC$ $^\text{b}$   && $-$32.420 993 7 && $-$59.205 089 2 && -94.028 041 9 && $-$187.885 492 \\
       $E_\DC^{(3)} = E_\text{nr}+\alpha^2 \varepsilon^\text{FW}_\DC + \alpha^3\varepsilon^{++}_\text{CC}$ $^\text{b}$  && $-$32.421 001 4 && $-$59.205 109 2 && $-$94.028 083 0 && $-$187.885 611  \\
       \hline \hline
    \end{tabular}
    ~\\[0.5cm]
    \begin{tabular}{{@{}l  lr lr lr@{}}}
    \hline \hline
       $Z$  && \multicolumn{1}{c}{18} &\hspace{0.5cm}& \multicolumn{1}{c}{22} &\hspace{0.5cm}& \multicolumn{1}{c}{26} \\ \hline
       $E_\mathrm{DC}^\mathrm{proj}$  && $-$314.246 103 && $-$473.436 340 && $-$665.885 818 \\
       $E_\mathrm{DC,Hy-CI}^\mathrm{proj}$ \cite{bylickiRelativisticHylleraasConfigurationinteraction2008}  && $−$314.246 117 && $−$473.436 350  && $−$665.885 871   \\
       $E_\mathrm{MCDF}^\mathrm{proj}$ \cite{parpiaAccurateDiracCoulombEnergies1990}  && $−$314.246 0 \hspace{0.2cm} && $−$473.436 2 \hspace{0.2cm} && $−$665.885 6  $\hspace{0.2cm}$\\
       $E_\text{nr}$  $^\text{b}$ && $-$312.907 186 &$\hspace{0.5cm}$& $-$470.407 273 &$\hspace{0.5cm}$& $-$659.907 330 \\
       $E_\DC^{(2)} = E_\text{nr}+\alpha^2 \varepsilon^\text{FW}_\DC$ $^\text{b}$   && $-$314.233 989 && $-$473.396 895 && $-$665.776 191 \\
       $E_\DC^{(3)} = E_\text{nr}+\alpha^2 \varepsilon^\text{FW}_\DC + \alpha^3\varepsilon^{++}_\text{CC}$ $^\text{b}$  && $-$314.234 250 && $-$473.397 382 && $-$665.777 007 \\
       \hline \hline
    \end{tabular}
    \begin{flushleft}
      $^\text{a}$ %        
        The non-relativistic energy, $E_\text{nr}$, and the expectation values used to evaluate the $\varepsilon^\text{FW}_\DC$ and $\varepsilon^{++}_\text{CC}$ correction terms were taken from Ref.~\cite{drakeHighPrecisionCalculations2006}. \\
      $^\text{b}$ %        
        The non-relativistic energy, $E_\text{nr}$, and the expectation values used to evaluate the $\varepsilon^\text{FW}_\DC$ and $\varepsilon^{++}_\text{CC}$ correction terms were       
        directly computed with the ECG basis set. 
    \end{flushleft}
\end{table}

%\clearpage
%\bibliography{book,ComplexScaling,Dirac-Coulomb,KineticBalance,LowerBound,MathAndConstants,NonAdiabatic,H2,programs,Relativistic,relatquantum,ExplicitGauss,ExplicitExponential,He,Hthreep,qed,QED,experiments,theses}

\end{document}